\documentclass[aps,preprint,nofootinbib,eqsecnum]{revtex4}%
\usepackage{amsmath}
\usepackage{graphicx}
\usepackage{amsfonts}
\usepackage{amssymb}%
\providecommand{\U}[1]{\protect\rule{.1in}{.1in}}
\providecommand{\U}[1]{\protect\rule{.1in}{.1in}}
\providecommand{\U}[1]{\protect\rule{.1in}{.1in}}
\providecommand{\U}[1]{\protect\rule{.1in}{.1in}}
\providecommand{\U}[1]{\protect\rule{.1in}{.1in}}

\begin{document}

\preprint{{\leftline {USC-09/HEP-B5 \hfill CERN-PH-TH/2010-188}}}
\title[Super Yang-Mills theory in 10+2 dimensions]{Super Yang-Mills Theory
in 10+2 Dimensions, \\
The 2T-physics Source for $\mathcal{N}$=4 SYM and M(atrix) Theory}
\thanks{This work was partially supported by the US Department of Energy,
grant number DE-FG03-84ER40168.\\
\bigskip \bigskip }
\author{Itzhak Bars and Yueh-Cheng Kuo\bigskip }
\affiliation{Department of Physics and Astronomy, University of Southern California, Los
Angeles, CA 90089-0484, USA\\
{\rm and}\\
Theory Division, Physics Department, CERN, 1211 Geneva 23, Switzerland\\
\bigskip }

\begin{abstract}
In this paper we construct super Yang-Mills theory in 10+2 dimensions
(SYM$_{10+2}^{1}$), a number of dimensions that was not reached before in a unitary
supersymmetric field theory, and show that this is the 2T-physics source of some cherished lower dimensional field theories. The much studied conformally exact $\mathcal{N}$=4
SuperYang-Mills field theory in 3+1 dimensions (SYM$_{3+1}^{4}$) is known to
be a compactified version of $\mathcal{N}$=1 SYM in 9+1 dimensions
(SYM$_{9+1}^{1}$), while M(atrix) theory is obtained by compactifications of
the 9+1 theory to 0 dimensions (also 0+1 and others). We show that there is a
deeper origin of these theories in two higher dimensions as they emerge from
the new SYM$_{10+2}^{1}$ theory with two times. Pursuing various alternatives
of gauge choices, solving kinematic equations and/or dimensional reductions of
the 10+2 theory, we suggest a web of connections that include those mentioned
above and a host of new theories that relate 2T-physics and 1T-physics field
theories, all of which have the 10+2 theory as the parent. In addition to
establishing the higher spacetime underpinnings of these theories, a side benefit
could be that in principle our approach can be used to develop new computational techniques.

\end{abstract}
\maketitle
\tableofcontents

\newpage

\section{The action, and summary of results}

The rules of 2T field theory in flat space \cite{2tstandardM}%
-\cite{emergentfieldth2} and in curved space \cite{2tGravity}-\cite{2tScalars}
in $d+2$ dimensions are well established \cite{phaseSpace}. They are derived
from an underlying Sp$\left(  2,R\right)  $ gauge symmetry principle in phase
space at the worldline level, which in turn leads to a \textit{ghost free
unitary 2T field theory} thanks to new gauge symmetries and kinematic
constraints that follow from the 2T field theory action. These gauge symmetry
principles provide the fundamental answer to the question of how to construct
a physical theory in a spacetime with two timelike dimensions and still avoid
unitarity and causality problems. 2T field theory constructed with these rules
is compatible with conventional 1T field theory, but beyond this consistency,
2T-physics makes predictions that are missed in 1T physics systematically.
These include dualities and hidden symmetries \cite{emergentfieldth1}%
\cite{emergentfieldth2} and restrictions on the interactions of scalar fields
in usual relativistic 1T field theory \cite{2tstandardM}\cite{2tGravDetails}%
\cite{2tCosmo}\cite{2tScalars}. The new predictions, in the contexts of
classical mechanics, quantum mechanics or field theory, are all consistent
with known phenomenology at all scales of physics explored so far
\cite{phaseSpace}.

Following these rules the Lagrangian for the vector supermultiplet $(A_{M}%
^{a},\lambda_{A}^{a})$ and its coupling to gravity fields $\left(
G_{MN},\Omega,W\right)  $ in special $d+2=5,6,8,12$ dimensions is constructed
uniquely as follows
\begin{equation}
S=S_{SYM}+S_{Gravity}+\cdots\label{action1}%
\end{equation}
where the part that concerns this paper is
\begin{equation}
S_{SYM}=K\int d^{d+2}X\sqrt{-G}\delta\left(  W\left(  X\right)  \right)
\left\{  -\frac{1}{4g_{YM}^{2}}\Omega^{2\frac{d-4}{d-2}}F_{MN}^{a}F_{a}%
^{MN}+\frac{i}{2}\left[  \overline{\lambda}^{a}V\bar{D}\lambda^{a}%
+\overline{\lambda}^{a}\overleftarrow{D}\overline{V}\lambda^{a}\right]
\right\}  .\label{action}%
\end{equation}
Note the unusual but important factor $\delta\left(  W\left(  X\right)
\right)  $ in the volume element which is essential in 2T field theory, where
the $W\left(  X\right)  $ field, along with the dilaton field $\Omega\left(
X\right)  $ are members of the \textquotedblleft gravity
triplet\textquotedblright\ $\left(  G_{MN},\Omega,W\right)  $. In $L_{SYM},$
the Yang-Mills field $A_{M}^{a}\left(  X\right)  $ is a vector in $d+2$
dimensions $X^{M}$ with $M=0^{\prime},1^{\prime},0,1,..,(d-1),$ with two
timelike components, $0,0^{\prime}$, while $\lambda_{A}^{a}\left(  X\right)  $
is a Weyl or Majorana spinor of SO($d,2$) (with 32 real components for
$d+2=12$ labelled by $A=1,2,...,32).$ Both $A_{M}^{a},\lambda_{A}^{a}$ are in
the adjoint representation of a the Yang-Mills gauge group G with
$a=1,2,\cdots,\dim\left(  adj\right)  .$ The scalar fields $\Omega$ and $W$,
together with the metric $G_{MN},$ are necessary to build up the action
$S_{Gravity}$ for 2T-Gravity as recently constructed \cite{2tGravity} and
analyzed in great detail \cite{2tGravDetails}\cite{2tCosmo},%
\begin{equation}
S_{Gravity}=K\int d^{d+2}X\sqrt{G}\left[
\begin{array}
[c]{c}%
\delta\left(  W\right)  \left\{  a_{d}\Omega^{2}R\left(  G\right)  +\frac
{1}{2}\partial\Omega\cdot\partial\Omega-V\left(  \Omega\right)  \right\} \\
+\delta^{\prime}\left(  W\right)  \left\{  a_{d}\Omega^{2}\left(  4-\nabla
^{2}W\right)  +a_{d}\partial W\cdot\partial\Omega^{2}\right\}
\end{array}
\right] \label{sugra}%
\end{equation}

Our spinor conventions and gamma matrices $\Gamma^{i},\bar{\Gamma}^{i}$ for
SO$\left(  10,2\right)  $ are given in footnote (\ref{gammas}) and in great
detail in the appendix of ref.\cite{susy2tN1}. The symbols $V\equiv\Gamma
^{M}V_{M}=\Gamma^{i}V_{i}$ and $\bar{V}\equiv\bar{\Gamma}^{M}V_{M}=\bar
{\Gamma}^{i}V_{i},$ with gamma matrices $\Gamma^{M}=\Gamma^{i}E_{i}^{M}$ that
appear in $L_{SYM},$ contains the fields $V_{M}\left(  X\right)  \equiv
\frac{1}{2}\partial_{M}W$ and the vielbein $E_{M}^{i}$ explained in
Eq.(\ref{geom}) below.

The supersymmetric completion of $S_{Gravity}$ to supergravity symbolized by
\textquotedblleft$+\cdots$\textquotedblright\ in Eq.(\ref{action1}) has been
obtained in 4+2 dimensions \cite{2tsugra}, but it remains incomplete in 10+2
dimensions at this stage. Therefore, in this paper the fields $\Omega
,W,G_{MN}$ will be treated as if they are non-dynamical backgrounds for the
purpose of supersymmetry transformations. So, $W,\Omega,G_{MN}$ \textit{will
not transform} under SUSY. Except for \textit{kinematical} equations (not
dynamical ones, see below) of the background fields $\Omega,W,G_{MN}$ given in
Eq.(\ref{geom}), that follow from varying $S_{Gravity}~$\cite{2tGravity}%
\cite{2tGravDetails}, without involving $S_{SYM}$, or the missing terms
\textquotedblleft\ $\cdots$\textquotedblright, the gravitational sector
$\left(  S_{Gravity}+\cdots\right)  $ will not play a further role in
determining the supersymmetry or other structural properties of $S_{SYM}$.
This is sufficient to construct and interpret SYM in $d+2=5,6,8,12.$

In 2T field theory, the gauge field $F_{MN}^{a}=\partial_{M}A_{N}^{a}%
-\partial_{N}A_{M}^{a}+f^{abc}A_{M}^{b}A_{N}^{c}$ must couple to $\Omega$ and
$G_{MN}$ in the action in the form $-\frac{1}{4g^{2}}\Omega^{2\frac{d-4}{d-2}%
}F_{MN}^{a}F_{PQ}^{a}G^{MP}G^{NQ}$ in $d+2$ dimensions, but note that the
dilaton factor disappears in $4+2$ dimensions. The covariant derivative of the
spinor $\bar{D}\lambda_{A}^{a}$ contains the SO$\left(  d,2\right)  $ spin
connection $\omega_{M}^{ij}$ and the inverse vielbein $E_{k}^{M},$ in addition
to the Yang-Mills field $A_{M}^{a}.$ Note also the notation $\overline
{\lambda}^{a}\overleftarrow{D}\equiv\overline{D_{M}\lambda}^{a}\Gamma^{M}.$
\begin{equation}
\bar{D}\lambda^{a}\equiv\bar{\Gamma}^{M}D_{M}\lambda^{a}=\bar{\Gamma}^{k}%
E_{k}^{M}\left(  \partial_{M}\lambda^{a}+\frac{1}{4}\omega_{M}^{ij}\Gamma
_{ij}\lambda^{a}+f^{abc}A_{M}^{b}\lambda^{c}\right)  .
\end{equation}

The variation of the action with respect to each field produces terms
proportional to $\delta\left(  W\right)  ,\delta^{\prime}\left(  W\right)  $
and $\delta^{\prime\prime}\left(  W\right)  .$ Each coefficient must vanish
since these are linearly independent distributions. The terms proportional to
$\delta^{\prime}\left(  W\right)  $ and $\delta^{\prime\prime}\left(
W\right)  $ are the \textquotedblleft kinematic\textquotedblright\ equations
while the terms proportional to $\delta\left(  W\right)  $ are the
\textquotedblleft dynamical\textquotedblright\ equations. The dynamical
equations for each field contain interactions, but the kinematical ones do not
(except for those that enter through gauge invariant derivatives, but those
interactions vanish in special gauge choices). So, in addition to the usual
geometric relations (as found in standard general relativity textbooks) among
the vielbein $E_{M}^{i},$ metric $G_{MN}=E_{M}^{i}E_{N}^{j}\eta_{ij}$, spin
connection $\omega_{M}^{ij}$, and affine connection $\Gamma_{MN}^{P}$ (not to
be confused with gamma matrices), the following \textit{additional kinematic
equations} among geometrical quantities in 2T gravity in $d+2$ dimensions must
also be imposed on the gravitational fields $W,\Omega,G_{MN},E_{M}^{i}%
,\omega_{M}^{ij},\Gamma_{MN}^{P}.$ It is important to emphasize that these
\textquotedblleft kinematic\textquotedblright\ equations follow from varying
the action for $S_{Gravity}$; they are not imposed from outside as additional
constraints.
\begin{equation}%
\begin{array}
[c]{c}%
V_{M}=\frac{1}{2}\partial_{M}W,\;V^{M}=G^{MN}V_{N},\;V^{i}=V^{M}E_{M}^{i},\;\\
W=V^{i}V_{i}=G^{MN}V_{M}V_{N}=\frac{1}{2}V^{M}\partial_{M}W,\;\\
G_{MN}=\nabla_{M}V_{N}=\frac{1}{2}\left(  \partial_{M}\partial_{N}%
W-\Gamma_{MN}^{P}\partial_{P}W\right)  ,\;\\
\;E_{M}^{i}=D_{M}V^{i}=\partial_{M}V^{i}+\omega_{M}^{ij}V_{j},\\
\left(  V^{M}\partial_{M}+\frac{d-2}{2}\right)  \Omega=0.
\end{array}
\label{geom}%
\end{equation}
These equations can also be derived directly from the Sp$\left(  2,R\right)  $
gauge symmetry principle that underlies 2T-physics at the worldline level in a
curved background that includes $G_{MN}$ \cite{2tGravity}. The significance of
these kinematical equations is to restrict the degrees of freedom to gauge
invariant sectors of the underlying Sp$(2,R)$ gauge symmetry in curved
backgrounds \cite{2tGravity}\cite{2tGravDetails}\cite{2tScalars}. Through
these equations, the scalar field $W$ determines some of the properties of
geometrical quantities such as $G_{MN},E_{M}^{i},$ etc. Geometrically, these
are \textit{homothety conditions} on the metric $G_{MN}$ and other fields
\cite{2tGravity}\cite{2tGravDetails}\cite{2tScalars}.

These equations are solved by flat spacetime in $d+2$ dimensions as well as by
the most general curved spacetime in $d$ dimensions (less one time and one
space dimension) as embedded in $d+2$ dimensions\footnote{Flat space in $d+2$
dimensions obeys Eq.(\ref{geom}) with $W_{flat}=X\cdot X=\eta_{MN}X^{M}X^{N},$
$V_{M}^{flat}=X_{M}$, $G_{MN}^{flat}=\eta_{MN},$ $\left(  \Gamma_{MN}%
^{p}\right)  _{flat}=0=\left(  \omega_{M}^{ij}\right)  _{flat}$ and
$\Omega_{flat}=\left(  c\cdot X\right)  ^{1-d/2}$ with a constant $c_{M}.$ A
curved metric that satisfies Eq.(\ref{geom}) can be taken in the form
$G_{MN}=\eta_{MN}+h_{MN}\left(  X\right)  ,$ still with $W=\eta_{MN}X^{M}%
X^{N}$, $V^{M}=X^{M}$, $V_{M}=\eta_{MN}X^{N}$, but with $X^{M}h_{MN}=0,$
$X\cdot\partial h_{MN}=0.$ Other forms of solutions of Eq.(\ref{geom}) in
curved space, that are more convenient to describe the conformal shadow, are
found in \cite{2tGravity}\cite{2tGravDetails}; see also the text and
appendix-\ref{flat} in this paper. The solution for all such $G_{MN}$
corresponds to the most general unrestricted background metric $g_{\mu\nu
}\left(  x\right)  $ in $d$ dimensions $x^{\mu}$ \cite{2tbacgrounds} plus
\textquotedblleft prolongations\textquotedblright\ in the extra dimensions
\cite{2tGravDetails}. Depending on the shadow (see footnote \ref{shadows}),
the prolongations are determined by $g_{\mu\nu}\left(  x\right)  $ or are
gauge freedom; they are not dynamical Kaluza-Klein modes. \label{backgrounds}%
}. In the general solution there are no Kaluza-Klein type degrees of freedom
that connect the \textquotedblleft shadow\textquotedblright\ in $d$ dimensions
and the \textquotedblleft substance\textquotedblright\ in $d+2$ dimensions.
There are prolongations of the shadow \cite{2tGravDetails} that extend into
$d+2$ dimensions, but they are constructed from the degrees of freedom of the
shadow within $d$ dimensions, and these prolongations do not play any role in
determining the 1T-physics observed within the shadow.

In this paper we will establish the properties of this theory as summarized in
Fig.1. In section II we will discuss the central box at the top of Fig.1 for
SYM$_{d+2}^{1}$ for $d+2=12,8,6,5$ and argue that SUSY holds thanks to the
following two essential properties%

%TCIMACRO{\FRAME{dtbpFU}{4.9813in}{3.7983in}{0pt}{\Qcb{Fig.1 - SYM$_{10+2}^{1}$
%is the parent of SYM$_{9+1}^{1},$ SYM$_{4+2}^{4},$ SYM$_{3+1}^{4},$ and
%M(atrix) theories.}}{\Qlb{fig_12Dsym}}{fig_sym12d.wmf}%
%width 4.9813in;  height 3.7983in;  depth 0pt;  original-width 9.1757in;
%original-height 6.9842in;  cropleft "0";  croptop "1";  cropright "1";
%cropbottom "0";  filename 'fig_sym12D.wmf';file-properties "XNPEU";}} }%
%BeginExpansion
%\begin{center}
%\includegraphics[
%height=3.7983in,
%width=4.9813in
%]%
%{fig_sym12D.eps}%
%\\
%Fig.1 - SYM$_{10+2}^{1}$ is the parent of SYM$_{9+1}^{1},$ SYM$_{4+2}^{4},$
%SYM$_{3+1}^{4},$ and M(atrix) theories.
%\label{fig_12Dsym}%
%\end{center}
%%EndExpansion

\begin{center}
\includegraphics[
height=3.5in, %4.4339in,
width=4.5in %5.8141in
]%
{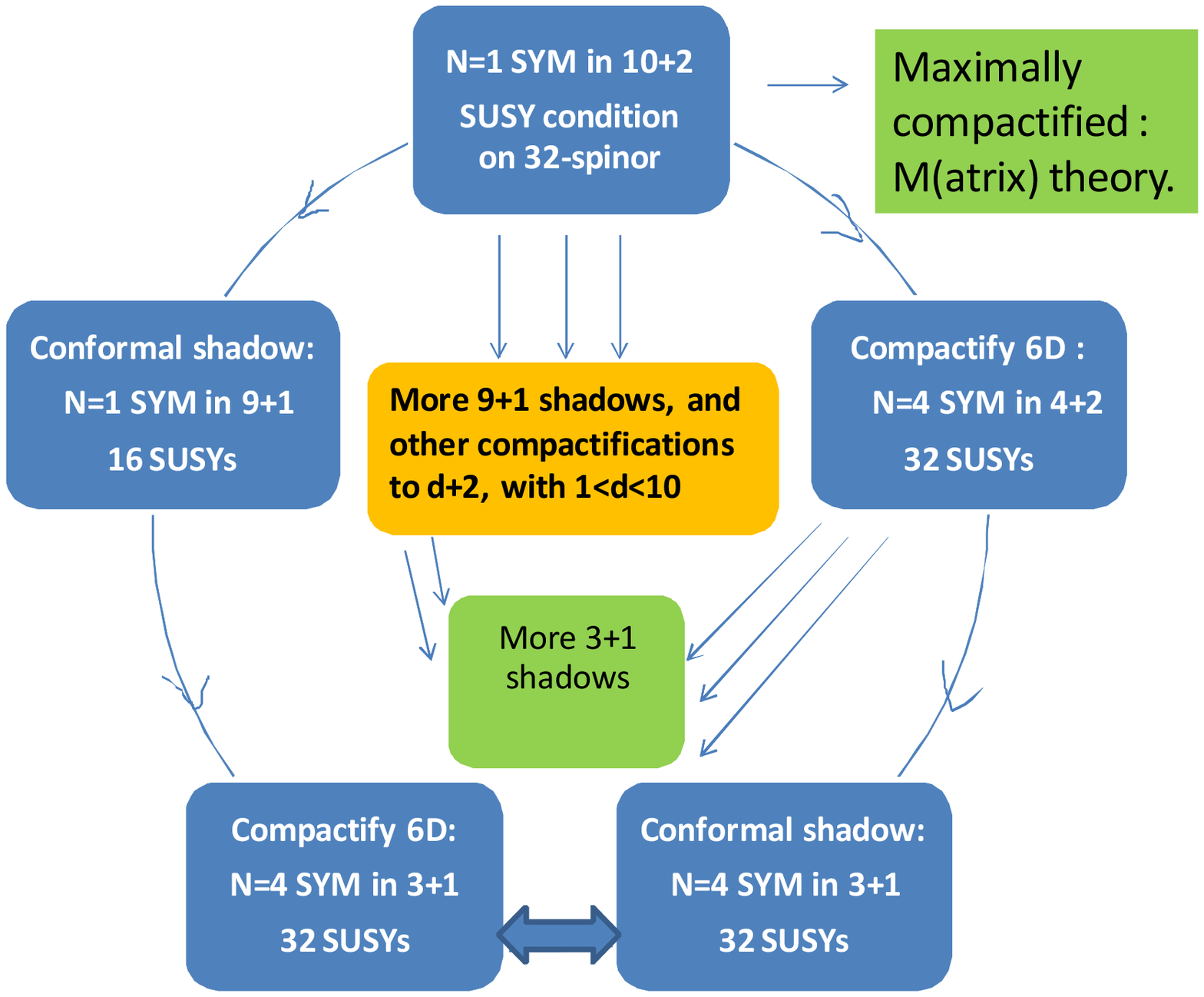}%
\\
Fig.1 - SYM$_{10+2}^{1}$ is the parent of SYM$_{9+1}^{1},$ SYM$_{4+2}^{4},$
SYM$_{3+1}^{4},$ and M(atrix) theories.
\label{fig_12Dsym}%
\end{center}

\begin{description}
\item[1-] The first ingredient is the following special identity for the gamma
matrices of SO$\left(  d,2\right)  ,~\Gamma^{ij}=\frac{1}{2}\left(  \Gamma
^{i}\Gamma^{j}-\Gamma^{j}\Gamma^{i}\right)  ,$
\begin{equation}
\left(  \Gamma^{ik}\right)  _{(AB}\left(  \Gamma_{k}^{\text{ \ }j}\right)
_{C)D}+\left(  \Gamma^{jk}\right)  _{(AB}\left(  \Gamma_{k}^{\text{ \ }%
i}\right)  _{C)D}=\frac{2\eta^{ij}}{d+2}\left(  \Gamma^{kl}\right)
_{(AB}\left(  \Gamma_{lk}\right)  _{C)D}.\label{gammaIds}%
\end{equation}
We derived this property and showed that it is satisfied only for
$d+2=12,8,6,5.$ Here the SO$\left(  d,2\right)  $ spinor indices are
symmetrized as implied by the parenthesis $\left(  ABC\right)  .$

\item[2- ] In addition, the local SUSY parameter $\varepsilon_{A}\left(
X\right)  $ must obey the following differential condition in the presence of
the curved spacetime backgrounds $G_{MN},\Omega,W~$consistently with
Eq.(\ref{geom})
\begin{equation}
\left\{  -\frac{d-4}{d-2}\left(  \bar{\Gamma}^{PQN}\Gamma^{M}\varepsilon
\right)  _{A}V_{N}\partial_{M}\ln\Omega+\left(  \bar{\Gamma}^{M}\Gamma
^{PQN}D_{M}\varepsilon\right)  _{A}V_{N}=V^{P}U_{A}^{Q}-V^{Q}U_{A}%
^{Q}\right\}  _{W=0}\label{SUSYcond}%
\end{equation}

\end{description}

The $U_{A}^{Q}\left(  X\right)  $ in Eq.(\ref{SUSYcond}) is an arbitrary
vector-spinor. Solutions of this equation will be discussed at the end of
section II and in Appendix (\ref{Scondition}). We emphasize that the SUSY
condition (\ref{SUSYcond}) arises because the background fields do not
transform under SUSY. In dynamical 2T supergravity in $d+2$ dimensions
\cite{2tsugra}, where $G_{MN},\Omega,W$ and the gravitino $\Psi_{MA}$ also
undergo SUSY transformations, the transformation of the gravitino field,
$\delta_{\varepsilon}\Psi_{MA}=D_{M}\varepsilon_{A}+\cdots,$ will cancel at
least the $D_{M}\varepsilon$ part of this expression thus removing or altering
this condition on $\varepsilon_{A}\left(  X\right)  $.

In section IIIA we will outline the derivation of SYM$_{9+1}^{1}$ as the
\textquotedblleft conformal shadow\textquotedblright\ of SYM$_{10+2}^{1}$
taken in a \textit{flat background. }The conformal shadow\footnote{Nontrivial
examples of 1T shadows from 2T-physics has been given in classical or quantum
mechanics and in field theory. For the simplest flat $d+2$ dimensional
background described in footnote (\ref{backgrounds}) see the figures in
\cite{phaseSpace} and the corresponding formulas for the shadows summarized in
tables I, II and III in \cite{emergentfieldth1}, which includes examples of
shadows in field theory (see also \cite{emergentfieldth2}). The conformal
shadow is the one most familiar to particle physicists. Therefore it has
featured as an explicit example in many old \cite{Dirac}-\cite{marnelius} and
recent discussions \cite{weinberg} in addition to discussions in many papers
by the current authors \cite{phaseSpace}, to help absorb some of the 1T
physical content in 2T-physics. The richness of the predictions of 2T-physics,
which is missed in the conventional formulation of 1T-physics, is in the
presence of the many other shadows such as those summarized in
\cite{phaseSpace}\cite{rivelles}. \label{shadows}} is arrived at as a
combination of a special gauge choice of 2T gauge symmetries and the solution
of kinematic constraints on $A_{M}^{a},\lambda_{A}^{a}$ derived from
$L_{SYM}.$ It is well known that SYM$_{9+1}^{1}$ has only 16 supersymmetries
while its compactification to SYM$_{3+1}^{4}$ depicted in Fig.1 results in the
intensely studied $\mathcal{N}=4 $ SYM theory in $3+1$ dimensions with 32
supersymmetries within the supergroup SU$\left(  2,2|4\right)  $.

In section IIIB we will obtain SYM$_{4+2}^{4}$ as a straightforward
compactification of the Lagrangian $L_{SYM}$ into a 2T field theory in flat
4+2 dimensions with 32 supersymmetries. The unique SYM$_{4+2}^{4}$ theory was
previously constructed by us by direct 2T SUSY methods in $4+2$ dimensions
\cite{susy2tN2N4}. We had previously argued that the \textquotedblleft
conformal shadow\textquotedblright\ of the unique SYM$_{4+2}^{4}$ is the
$\mathcal{N}=4$ super Yang-Mills theory SYM$_{3+1}^{4}.$ So, the left and
right sides in Fig.1 arrive at the same special SYM$_{3+1}^{4}$ with 32
supersymmetries via different routes that display the consistency and some of
the properties of the parent SYM$_{10+2}^{1}$ theory$.$ We expect that the
parent SYM$_{10+2}^{1}$ theory, and its compactification to SYM$_{4+2}^{4}$,
together with the various shadows that are related by dualities
\cite{emergentfieldth1}\cite{emergentfieldth2}, would add new tools and shed
new light on the intensely studied SYM$_{3+1}^{4}$ theory.

In section IV we will derive the 2T version of M(atrix) theory (top right in
Fig.1) by dimensionally reducing 9 or 9+1 dimensions, leaving behind a 2T
M(atrix) theory with either 1+2 or 1+1 dimensions with certain gauge
symmetries. The conformal shadows of this 2T M(atrix) theory in 1+1 or 1+2
dimensions yields the familiar versions of 1T M(atrix) theory that describe
(-1)-branes, 0-branes, or more generally p-branes \cite{AreaDiff}%
-\cite{matrixThJ4}.

There are many other routes of deriving supersymmetric theories from
SYM$_{10+2}^{1}$ either by exploring the many shadows of 2T-physics other than
the conformal shadow or by considering other compactifications, as well as a
variety of backgrounds $G_{MN},\Omega,W$. These are indicated schematically in
Fig.1. We will make only brief comments on these possibilities.

\section{SUSY condition}

The SUSY transformation of the dynamical fields $A_{M}^{a},\lambda_{A}^{a}$ is
similar to the one we discussed previously in 4 +2 dimensions \cite{susy2tN1}%
\cite{susy2tN2N4} but here it is modified for $d+2=5,6,8,12$ dimensions and
the presence of the background fields, $G_{MN},\Omega,W,\omega_{M}^{ij}%
,E_{M}^{i},$ which were absent in \cite{susy2tN1}\cite{susy2tN2N4}:
\begin{equation}
\delta_{\varepsilon}\lambda_{A}^{a}=\frac{i}{g_{YM}}\Omega^{\frac{d-4}{d-2}%
}F_{MN}^{a}\left(  \Gamma^{MN}\varepsilon\right)  _{A},\;\;\delta
_{\varepsilon}A_{M}^{a}=\Omega^{-\frac{d-4}{d-2}}\left[  -2\bar{\varepsilon
}\Gamma_{M}\bar{V}\lambda^{a}+W\bar{\varepsilon}\Gamma_{MN}D^{N}\lambda
^{a}\right]  +h.c..\label{delAL}%
\end{equation}
It takes some effort to verify that the action (\ref{action1}) is invariant
$\delta_{\varepsilon}S=0$ under (\ref{delAL},\ref{kineps}), \textit{without
varying the background fields}. After taking into account the kinematic
properties of the curved background in Eq.(\ref{geom}), which is discussed in
detail in Sec.IIIB of \cite{2tGravDetails}, one finds that the algebraic
manipulations to verify SUSY are completely parallel to those in flat $4+2$
dimensions given in \cite{susy2tN1}, and the proof proceeds by formally
replacing derivatives by covariant derivatives, etc., in the presence of the
backgrounds. So, we will only state that indeed we find $\delta_{\varepsilon
}S=0$ by following the steps of the computation in \cite{susy2tN1}. The
crucial equations (\ref{gammaIds},\ref{SUSYcond}) are the only new ingredients
necessary to show the symmetry of the action in $d+2$ dimensions, with
$d+2=12,8,6,5$. In particular, the SUSY condition (\ref{SUSYcond}) is new (but
trivially satisfied for $d+2=6$).

An alternative proof of supersymmetry is to show that there is a conserved
SUSY current, $0=\partial_{M}\left(  \bar{\varepsilon}^{A}J_{A}^{M}\right)
=\left(  D_{M}\bar{\varepsilon}^{A}\right)  J_{A}^{M}+\bar{\varepsilon}%
^{A}D_{M}J_{A}^{M},$ where $D_{M}$ includes the background spin connection.
The current derived by using Eqs.(\ref{delAL}) and Noether's theorem is
\begin{equation}
\bar{\varepsilon}J^{M}=\delta\left(  W\right)  \sqrt{G}\Omega^{\frac{d-4}%
{d-2}}F_{PQ}^{a}V_{N}~\bar{\varepsilon}\left(  \Gamma^{PQN}\bar{\Gamma}%
^{M}\right)  \lambda^{a}.\label{current}%
\end{equation}
To show the conservation, $\partial_{M}\left(  \bar{\varepsilon}J^{M}\right)
=0,$ we must use the equations of motion derived from the action. The
\textquotedblleft kinematic equations\textquotedblright, namely those that
come from terms proportional to $\delta^{\prime}\left(  W\right)  $ in the
variation of the action are\footnote{In proving the conservation of the
current we must also include a kinematic condition on the SUSY parameter
\begin{equation}
V\cdot D\varepsilon_{A}\equiv V^{M}\left(  \partial_{M}\varepsilon_{A}%
+\frac{1}{4}\omega_{M}^{ij}\left(  \Gamma_{ij}\varepsilon\right)  _{A}\right)
=0.\label{kineps}%
\end{equation}
This is required since in this computation all fields are on shell constrained
by kinematic equations (as a result of equations of motion), whose
significance is the imposition of Sp$\left(  2,R\right)  $ gauge invariance.}
\begin{equation}
V^{M}F_{MN}^{a}=0,\;\;\left(  V\cdot D+\frac{d}{2}\right)  \lambda_{A}%
^{a}=0,\label{kin}%
\end{equation}
in addition to those listed in Eqs.(\ref{geom},\ref{kineps}). The
\textquotedblleft dynamical equations\textquotedblright, namely those that
come from terms proportional to $\delta\left(  W\right)  $ in the variation of
the action, are\footnote{Identifying the dynamical/kinematical equations from
a variation of the action that has the form $\delta S\sim\int\delta\Phi\left[
\alpha\left(  X\right)  \delta\left(  W\right)  +\beta\left(  X\right)
\delta^{\prime}\left(  W\right)  \right]  =0$ requires also a discussion of
gauge symmetry. For a recent discussion see section-V in \cite{2tGravDetails}
for how a gauge is chosen to arrive at the kinematic equation $\beta=0$ at all
$W\left(  X\right)  ,$ and the dynamical equation $\alpha=0$ at $W\left(
X\right)  =0,$ and how this relates to an underlying Sp$\left(  2,R\right)  $
symmetry.}
\begin{equation}
\left(  V\bar{D}\lambda^{a}\right)  _{A}=0,\;\;\hat{D}_{N}\left(
\Omega^{\frac{2\left(  d-4\right)  }{d-2}}F_{a}^{NM}\right)  =f_{abc}\left(
\overline{\lambda^{b}}\Gamma^{MN}\lambda^{c}\right)  V_{N}.\label{dyn}%
\end{equation}
These are required to be satisfied only on the shell $W=0.$

Additional properties of this current\footnote{This current can be modified by
additional inessential terms $\Delta J_{A}^{M}$ that are automatically
conserved $\partial_{M}\left(  \bar{\varepsilon}\Delta J^{M}\right)  =0$ on
their own, independent of dynamics. Such terms, that are analogous to the
automatically conserved terms in the "new improved" energy momentum tensor,
have the forms $\Delta J_{A}^{M}=\delta(W)\sqrt{G}V^{M}\xi_{A}$ or $\Delta
J_{A}^{M}=\delta^{\prime}(W)\sqrt{G}V^{M}\tilde{\xi}_{A},$ where the spinors
$\xi_{A},\tilde{\xi}_{A}$ may be functions of the fields and must satisfy
homogeneity conditions $\left(  V\cdot D+d\right)  \xi_{A}=0$ and $\left(
V\cdot D+d-2\right)  \tilde{\xi}_{A}=0$ that follow only from the kinematical
equations (\ref{geom},\ref{kineps},\ref{kin}) for all fields including the
backgrounds. An example is $\Delta J_{A}^{M}=\delta(W)\sqrt{G}V^{M}%
\Omega^{\frac{d-4}{d-2}}F_{PQ}^{a}\left(  \Gamma^{PQ}\lambda^{a}\right)
_{A}.$ The automatic conservation is verified by noting some simple kinematic
relations, such as $\partial_{M}\left(  \sqrt{G}V^{M}\right)  =\sqrt{G}%
\nabla_{M}V^{M}=\sqrt{G}\delta_{M}^{M}=\left(  d+2\right)  \sqrt{G}$ and
$V\cdot\partial\delta(W)=\delta^{\prime}(W)V\cdot\partial W=2W\delta^{\prime
}(W)=-2\delta\left(  W\right)  ,$ where $V\cdot\partial W=2W$ was used
(Eq.\ref{geom}) and similarly $V\cdot\partial\delta^{\prime}(W)=-4\delta
^{\prime}(W).$ Then with only the kinematics one verifies $\partial_{M}\left(
\bar{\varepsilon}\Delta J^{M}\right)  =0,$ \textit{independent of the
dynamical equations} (\ref{dyn}). \label{irrelevant}} include that it is
orthogonal to $V_{M},$ namely $\bar{\varepsilon}J^{M}V_{M}=0,$ proven by using
the kinematic equations $W=V\cdot V=0$ and $V^{M}F_{MN}^{a}=0$ in
(\ref{geom},\ref{kin}) and applying $W\delta\left(  W\right)  =0$. It can also
be verified that this current is invariant under the following local
symmetries shared by the action (see also footnote (\ref{specialized})) :
$\left(  1\right)  $ Under the 2T gauge transformation of the gauge field
\cite{2tstandardM}, $\delta_{\Lambda}A_{M}^{a}=Ws_{M}^{a}\left(  X\right)  ,$
the totally antisymmetric form $\delta(W)F_{[PQ}V_{N]}$ that occurs in the
current is invariant, and $\left(  2\right)  $ under the 2T gauge
transformation of the gaugino $\delta_{\kappa}\lambda_{A}^{a}=\left(
V\kappa_{1}^{a}\right)  _{A}+W\kappa_{2A}^{a}$ \cite{2tstandardM} with local
fermionic parameters $\kappa_{1A}^{a}\left(  X\right)  ,\kappa_{2A}^{a}\left(
X\right)  ,$ the expression for $\delta_{\kappa}\left(  \bar{\varepsilon}%
J^{M}\right)  $ vanishes modulo the irrelevant types of terms described in
footnote (\ref{irrelevant}), or contains terms proportional to the kinematic
equations (\ref{geom},\ref{kin}) which also vanish. So, although the fermionic
2T gauge transformation of the current $\delta_{\kappa}\left(  \bar
{\varepsilon}J^{M}\right)  $ is not strictly zero, it may be ignored in the
Sp$\left(  2,R\right)  $ gauge invariant sector, since it vanishes when only
the kinematic equations are put on shell, while the dynamic equations
(\ref{dyn}) are not imposed.

We emphasize the following crucial points in proving the conservation of the
current $\partial_{M}\left(  \bar{\varepsilon}^{A}J_{A}^{M}\right)  =0.$ After
using both the kinematic and dynamical equations of motion, the divergence of
the current can be brought to the form
\begin{equation}
\partial_{M}\left(  \bar{\varepsilon}J^{M}\right)  =\sqrt{G}\delta(W)\left\{
\begin{array}
[c]{c}%
2\Omega^{-\frac{d-4}{d-2}}f_{abc}V_{N}V^{P}\left(  \bar{\varepsilon}%
\Gamma^{QN}\lambda^{a}\right)  (\bar{\lambda}^{b}\Gamma_{QP}\lambda^{c})\\
+F_{PQ}V_{N}\left[
\begin{array}
[c]{c}%
-\partial_{M}\Omega^{\frac{d-4}{d-2}}\left(  \overline{\varepsilon}\Gamma
^{M}\bar{\Gamma}^{PQN}\lambda\right)  ~\\
+\left(  D_{M}\bar{\varepsilon}\right)  \Gamma^{PQN}\bar{\Gamma}^{M}\lambda
\end{array}
\right]
\end{array}
\right\}  .\label{div}%
\end{equation}
Now we use the special gamma matrix identity (\ref{gammaIds}) in $d+2$
dimensions (holds only for $d+2=12,8,6,5)$ to show that the first term in
(\ref{div}) vanishes%
\begin{equation}
f_{abc}V_{N}V^{P}\left(  \overline{\varepsilon}\Gamma^{QN}\lambda^{a}\right)
~\left(  \bar{\lambda}^{b}\Gamma_{QP}\lambda^{c}\right)  \delta(W)=\frac
{2}{d+2}f_{abc}\left(  \bar{\lambda}^{b}\Gamma_{kl}\lambda^{a}\right)  \left(
\bar{\varepsilon}\Gamma^{kl}\lambda^{c}\right)  W\delta(W)=0.\label{van}%
\end{equation}
The gamma matrix identity (\ref{gammaIds}) produces the second form in
(\ref{van}), but this identity alone is not sufficient to eliminate the first
term in (\ref{div}); we also need $W\delta(W)=0$ as in the last step of
(\ref{van}). The remaining expression in (\ref{div}) is in general
non-vanishing. However, if the SUSY parameter $\varepsilon^{A}\left(
X\right)  $ satisfies the condition (\ref{SUSYcond}) then this also vanishes
after using the kinematic equations, $V^{M}F_{MN}^{a}=0,\;W=V\cdot V$ and
$W\delta(W)=0,$ for any $U_{A}^{P}\left(  X\right)  $ in (\ref{SUSYcond}).

The discussion above provides an outline of the proof that SYM$_{10+2}^{1}$ is
indeed supersymmetric when $\varepsilon_{A}\left(  X\right)  $ satisfies the
SUSY condition (\ref{SUSYcond}). Now we want to show that there are solutions
for $\varepsilon_{A}$ that satisfy this condition. All solutions of
Eq.(\ref{SUSYcond}) are obtained in Appendix (\ref{Scondition}) by
concentrating on the conformal shadow. Below we display a specialized subclass
of simpler looking solutions that share some of the main features of the
general solution.

The simple class that obviously solves Eq.(\ref{SUSYcond}) is defined by
imposing stronger conditions on $\varepsilon_{A}\left(  X\right)  $ than
necessary, as follows
\begin{equation}
\left[  D_{M}\varepsilon\right]  _{W=0}\equiv\left[  \partial_{M}%
\varepsilon+\frac{1}{4}\omega_{M}^{ij}\Gamma_{ij}\varepsilon\right]
_{W=0}=0,\;\;\left[  \left(  \Gamma^{M}\varepsilon\right)  _{A}(\partial
_{M}\ln\Omega^{\frac{d-4}{d-2}})\right]  _{W=0}=0.\label{susyCond2}%
\end{equation}
In this case Eq.(\ref{SUSYcond}) is solved for $U_{A}^{Q}=0$ which, as
mentioned following (\ref{SUSYcond}), could be chosen arbitrarily. Note that
the second equation in (\ref{susyCond2}) becomes trivial in the case of
$d+2=6,$ so it constrains $\varepsilon_{A}$ only when $d+2=12,8,5$ but not
when $d+2=6.$ The first equation requires a covariantly constant spinor
$\left[  D_{M}\varepsilon\right]  _{W=0}=0$ in any of the curved backgrounds
that obey Eq.(\ref{geom})$.$ There are non-trivial backgrounds with
covariantly constant spinors\footnote{For a discussion of covariantly constant
spinors in non-trivial backgrounds see ref. \cite{GSW}, Eq.(15.1.3), and
related discussion in chapter 15. \label{ccspinors}} so it is of interest to
study those backgrounds that would be physically relevant in the applications
of SYM$_{10+2}^{1}.$

For a more explicit solution in $d+2$ dimensions we specialize further to the
flat background described in footnote (\ref{backgrounds}) which implies a
constant spinor $\partial_{M}\varepsilon_{A}=0$ since $\omega_{M}^{ij}=0.$ We
further take a special form for the dilaton $\Omega=\left(  c\cdot X\right)
^{1-d/2}$ with a constant vector $c_{M},$ to satisfy Eq.(\ref{geom}). Then
Eq.(\ref{susyCond2}) becomes $\left(  d-4\right)  c_{M}\left(  \Gamma
^{M}\varepsilon\right)  =0.$ By multiplying with another factor of
$c_{M}\Gamma^{M}$ we obtain the equation $\left(  d-4\right)  c^{2}%
\varepsilon_{A}=0.$ Evidently for $d+2=4$ the last equation puts no constraint
on $\varepsilon$ since it is trivially satisfied for any 4-component constant
complex spinor $\varepsilon_{A}$ (4 complex or 8 real fermionic parameters, so
8 supersymmetries which are part of SU$\left(  2,2|1\right)  $, with
SU$\left(  2,2\right)  =$SO$\left(  4,2\right)  $). However, in $d+2=12,8,5$
dimensions it requires a lightlike vector $c^{2}=0$ with $c_{M}\left(
\Gamma^{M}\varepsilon\right)  =0.$ This has solutions only when half of the
components of $\varepsilon_{A}$ vanish. Thus, for example, in $d+2=12$
dimensions, 16 out of the 32 real components of the constant SUSY parameter
must vanish. Hence SYM$_{10+2}^{1},$ when taken with a constant SUSY spinor in
a \textit{flat background} in $10+2$ dimensions, has at most 16 independent
real parameters in $\varepsilon_{A},$ and hence 16 non-trivial supersymmetries.

For the more general backgrounds that obey (\ref{geom}) as well as
Eq.(\ref{susyCond2}) with covariantly constant spinors $D_{M}\varepsilon=0,$ a
similar argument requires that $(\partial_{M}\ln\Omega^{\frac{d-4}{d-2}})$
should be a lightlike vector (when $d+2\neq6$) and therefore $\varepsilon
_{A}\left(  X\right)  $ still has at most 16 independent non-zero components
for $d+2=12$. However, since these are $X$-dependent, the number of
\textit{constant} parameters in the 16 non-zero components of the spinor
$\varepsilon_{A}\left(  X\right)  $ may exceed 16 in some backgrounds.

These results hold for the special class of solutions of Eq.(\ref{SUSYcond})
that follow from the stronger requirements in Eq.(\ref{susyCond2}). A similar
result holds also for the general solutions as discussed in Appendix
(\ref{Scondition}). However, when the background is curved, there are also
cases with 32 supersymmetries, as in the example of compactification from 10+2
to 4+2 dimensions shown on the right side of Fig.1 and treated in section
(\ref{gen4+2}).

\section{Shadows and compactifications}

In this section we will show that SYM$_{10+2}^{1}$ provides a higher
dimensional source and new perspectives for the popular SYM$_{9+1}^{1}$ and
SYM$_{3+1}^{4}$ that continue to be of intense interest in current research.
We will use usual techniques of dimensional reduction as well as techniques of
deriving shadows of 2T-physics \cite{2tstandardM}\cite{emergentfieldth1}%
\cite{emergentfieldth2}\cite{2tGravity}\cite{2tGravDetails} to obtain the
lower dimensional theories.

\subsection{Conformal shadow of SYM$_{10+2}^{1}$ gives SYM$_{9+1}^{1}$
\label{shadow1}}

We first briefly describe the result and then show how it is derived. We
choose a set of coordinates $X^{M}=\left(  w,u,x^{\mu}\right)  $ such that the
function $W\left(  X\right)  $ is simply $W\left(  X\right)  =w$ in terms of
the new coordinates. To see how such a basis can be chosen even in flat space
see Appendix \ref{flat}. Here we also explain how the general background
metric $ds^{2}=dX^{M}dX^{N}G_{MN}$ is brought to a basis that is convenient to
generate the conformal shadow as in \cite{2tGravity}\cite{2tGravDetails} while
imposing $w=0$ as required by the delta function $\delta\left(  W\left(
X\right)  \right)  $ in the action. In the set of coordinates $\left(
w,u,x^{\mu}\right)  $ we can solve all the kinematic constraints in
Eqs.(\ref{geom},\ref{kin},\ref{kineps}) for both the background and dynamical
fields. We will show that by a series of gauge choices and solving the
kinematic constraints we end up with the following shadow field configuration:
The original fields $A_{M}^{a},F_{MN}^{a},\lambda_{A}^{a}$ and $\Omega,G_{MN}$
in $d+2$ dimensions are then expressed in terms of the shadow fields at
$W\left(  X\right)  =w=0$ as functions of the remaining coordinates $u$ and
$x^{\mu}$ as follows
\begin{equation}
\;%
\begin{array}
[c]{ll}%
A_{M}^{a}\left(  X\right)  =\left\{
\begin{array}
[c]{l}%
A_{\mu}^{a}=A_{\mu}^{a}\left(  x\right)  ,\\
A_{w}=A_{u}=0,
\end{array}
\right.  \;\; & F_{MN}^{a}\left(  X\right)  =\left\{
\begin{array}
[c]{l}%
F_{\mu\nu}^{a}=F_{\mu\nu}^{a}\left(  x\right)  ,\\
F_{w\mu}^{a}=F_{u\mu}^{a}=F_{wu}^{a}=0,
\end{array}
\right. \\
\lambda_{A}^{a}\left(  X\right)  =\left(
\begin{array}
[c]{c}%
\lambda_{\alpha}\left(  x\right) \\
0
\end{array}
\right)  e^{\left(  d-1\right)  u}~~,\; &
\begin{array}
[c]{l}%
\Omega\left(  X\right)  =e^{\left(  d-2\right)  u}\phi\left(  x\right)  ,\\
G_{MN}\left(  X\right)  =\left\{
\begin{array}
[c]{l}%
G_{\mu\nu}=e^{-4u}g_{\mu\nu}\left(  x\right)  ,\;G_{wu}=-1,\\
G_{ww}=G_{w\mu}=G_{u\mu}=0.
\end{array}
\right.
\end{array}
\end{array}
\label{fieldShadows}%
\end{equation}
The shadow fields $A_{\mu}^{a}\left(  x\right)  ,\lambda_{\alpha}\left(
x\right)  $ form precisely the Yang-Mills supermultiplet in $d=10,6,4,3$
dimensions in a \textit{background} shadow spacetime described by $g_{\mu\nu
}\left(  x\right)  ,\phi\left(  x\right)  $. Note that there are no
Kaluza-Klein degrees of freedom since for example $A_{M}^{a}\left(  X\right)
\rightarrow A_{\mu}^{a}\left(  x\right)  ,$ and similarly for the other
fields. Having solved all the kinematic constraints (which amounts to imposing
Sp$\left(  2,R\right)  $ invariance), our original action in $d+2$ dimensions
can now be reduced to the conformal shadow action in $d$ dimensions that
includes gravity coupled to a conformally coupled dilaton $\phi$ (with the
wrong sign kinetic term) \cite{2tGravity}\cite{2tGravDetails}
\begin{equation}%
\begin{array}
[c]{c}%
S=S_{SYM}+\int d^{d}x\sqrt{-g}\left(  \frac{d-2}{8\left(  d-1\right)  }%
\phi^{2}R\left(  g\right)  +\frac{1}{2}\partial\phi\cdot\partial\phi\right)
,\\
S_{SYM}=\int d^{d}x\sqrt{-g}\left(  -\frac{1}{4g_{YM}^{2}}\phi^{2\frac
{d-4}{d-2}}F_{\mu\nu}^{a}F_{a}^{\mu\nu}+i\bar{\lambda}^{a}\gamma^{\mu}D_{\mu
}\lambda_{a}\right)  .
\end{array}
\label{actionShadow}%
\end{equation}
The shadow dilaton can be fixed to a constant $\phi\left(  x\right)
\rightarrow\phi_{0}$ by a Weyl transformation of all the fields\footnote{The
local scaling, known as the Weyl symmetry, is a natural outcome of 2T-gravity
\cite{2tGravDetails}. It arises as a remnant of the general coordinate
symmetry in the extra dimensions (there is no Weyl symmetry in the action in
$d+2$ dimensions). Using this remnant local symmetry, the negative norm
dilaton $\phi\left(  x\right)  $ can be removed as a degree of freedom, thus
insuring unitarity. Furthermore this Weyl gauge introduces Newton's
garvitational constant in the conformal shadow. Note that, even though
$\phi\left(  x\right)  $ can be set to a constant by a Weyl gauge, the
original dilaton field $\Omega\left(  X\right)  $ still depends on the extra
coordinate $u$, as given in Eq.(\ref{fieldShadows}). As discussed in
\cite{2tCosmo} other Weyl gauge choices for the dilaton, which also remove the
ghost $\phi\left(  x\right)  $, may be more convenient for certain useful
applications of the shadows concept.}. Then the part $S_{SYM}$ is recognized
as the action in $d=10,6,4,3$ dimensions for SYM$_{d}^{1}$ in a curved
background $g_{\mu\nu}\left(  x\right)  $ and a constant dilaton $\phi\left(
x\right)  =\phi_{0}$ with a dimensionful Yang-Mills coupling constant
(dimensionless only for $d=4$)
\begin{equation}
\hat{g}_{YM}=g_{YM}\phi_{0}^{-\frac{d-4}{d-2}}.
\end{equation}
Here the covariant derivative $D_{\mu}\lambda_{a}$ includes the Yang-Mills
gauge field as well the spin connection $\omega_{\mu}^{ab},$ and $\gamma^{\mu
}\equiv e_{a}^{\mu}\gamma^{a}$ includes the vielbein $e_{a}^{\mu}\left(
x\right)  $ associated with the general metric $g_{\mu\nu}\left(  x\right)  $.
Of course, the well known flat case in which $g_{\mu\nu}$ is fixed to the
Minkowski metric $\eta_{\mu\nu}$ and $\phi$ is fixed to a constant, is a
special case of the above.

The supersymmetry properties of the shadow action (\ref{actionShadow}) in
$d=3,4,6,10$ dimensions, in the presence of gravity and the dilaton
$\phi\left(  x\right)  $ (but not yet supergravity), follow from the SUSY
condition in $d+2$ dimensions (\ref{SUSYcond}), which is analyzed in detail in
Appendix (\ref{Scondition}), including the conserved SUSY current. From that
analysis we learn that this action is supersymmetric, without transforming
$g_{\mu\nu}\left(  x\right)  ,\phi\left(  x\right)  ,$ but transforming only
$A_{\mu}^{a}$ and $\lambda^{a}$ under SUSY, as follows
\begin{equation}
\delta_{\varepsilon}\lambda^{a}=\frac{i}{g_{YM}}\phi^{\frac{d-4}{d-2}}%
F_{\mu\nu}^{a}\gamma^{\mu\nu}\varepsilon,\;\;\delta_{\varepsilon}A_{\mu}%
^{a}=-2\phi^{-\frac{d-4}{d-2}}\bar{\varepsilon}\gamma_{\mu}\lambda^{a}+h.c.,
\end{equation}
provided the SO$\left(  d,1\right)  $ spinor SUSY parameter $\varepsilon
\left(  x\right)  $ satisfies the following conditions derived in Appendix
\ref{Scondition} (treating $g_{\mu\nu},\phi$ as backgrounds, indices
lowered/raised using $g_{\mu\nu}$)
\begin{equation}
D_{\mu}\varepsilon=\frac{1}{d}\gamma_{\mu}\left(  \bar{\gamma}\cdot
D\varepsilon\right)  \text{ and }\left(  d-4\right)  \bar{\gamma}^{\mu}D_{\mu
}\left(  \phi^{\frac{d}{d-2}}\varepsilon\right)  =0,\;\label{susys}%
\end{equation}
where $D_{\mu}\varepsilon_{\alpha}\left(  x\right)  =\partial_{\mu}%
\varepsilon_{\alpha}\left(  x\right)  +\frac{1}{4}\omega_{\mu}^{ab}\left(
x\right)  \left(  \gamma_{ab}\varepsilon\left(  x\right)  \right)  _{\alpha}$.
Note that the second equation is trivial for $d=4,$ so for $d=3,6,10$ there
are two constraints on $\varepsilon\left(  x\right)  ,$ but for $d=4$ only one
constraint. Here the spinors $\varepsilon$ or $\lambda$ have the following
numbers of components (this is half of the SO$(d,2)$ spinor, i.e.
$\varepsilon_{1}$ as indicated in Eq.(\ref{e1e2}))
\begin{equation}%
\begin{array}
[c]{l}%
d=3~:\;\text{the spinor of SO}\left(  2,1\right)  \text{ is real = a doublet
of SL}\left(  2,R\right)  ,\\
d=4~:\;\text{the Weyl spinor of SO}\left(  3,1\right)  \text{ = a complex
doublet of SL}\left(  2,C\right)  ,\\
d=6~:\;\text{the Weyl spinor of SO}\left(  5,1\right)  \text{, a complex
quartet.}\\
d=10:\;\text{the}~\text{Weyl-Majorana spinor of SO}\left(  9,1\right)
\text{~with 16 real components.}%
\end{array}
\end{equation}
We emphasize that the spinor $\varepsilon\left(  x\right)  $ is $x$-dependent,
and thus may contain more than one set of constant spinor parameters. This
number constant spinors, which determines the number of supersymmetries, will
depend on the background $g_{\mu\nu}\left(  x\right)  ,\phi\left(  x\right)  $
which in turn lead to the allowed solutions for $\varepsilon\left(  x\right)
$ in Eq.(\ref{susys}).

For example, consider the $d=4$ flat space background $g_{\mu\nu}=\eta_{\mu
\nu}$ with $\phi=\phi_{0}=$a constant. The solution of Eq.(\ref{susys}) is%
\begin{equation}
d=4,\text{ flat:\ }\varepsilon\left(  x\right)  =\varepsilon^{\left(
0\right)  }+x\cdot\gamma\varepsilon^{\left(  1\right)  },~\text{with
}\varepsilon^{\left(  0\right)  },\varepsilon^{\left(  1\right)  }\text{
constant SL}\left(  2,C\right)  \text{ doublets.}\label{d4susy}%
\end{equation}
In this case $\varepsilon^{\left(  0\right)  }$ corresponds to the usual
supersymmetry parameter while $\varepsilon^{\left(  1\right)  }$ corresponds
to the superconformal transformation parameter. The closure of these
transformations gives the global SU$\left(  2,2|1\right)  $ symmetry of
$\mathcal{N=}$1 super Yang-Mills theory in flat $d=4,$ which has 8
supersymmetries, namely the 8 real fermionic parameters in the two complex
SL$\left(  2,C\right)  $ doublets.

Repeating the same analysis for $d=10,6,3,$ still in the flat background, the
first equation has the same form as (\ref{d4susy}), but the second equation in
(\ref{susys}) eliminates $\varepsilon^{\left(  1\right)  }$, so that the
solution is
\begin{equation}
d=10,6,3,\text{ flat:\ ~}\varepsilon\left(  x\right)  =\varepsilon^{\left(
0\right)  }.
\end{equation}
Hence, for $d=10$ there are only 16, not 32 supersymmetries in a flat
background. However, in a curved background, in $d=10,$ the number could
decrease or increase. For example, it is well known that when 6 of the 10
dimensions of SYM$_{9+1}^{1}$ are compactified on a torus, the resulting
theory SYM$_{3+1}^{4}$ is $\mathcal{N=}$4 super Yang-Mills theory in flat
$d=4,$ which has SU$\left(  2,2|4\right)  $ symmetry, with 32 supersymmetries
(as shown on the left branch in Fig.1).

A similar analysis for various fixed non-flat backgrounds $g_{\mu\nu}\left(
x\right)  ,\phi\left(  x\right)  $ determines the number and nature of
supersymmetries. Whatever those are, they correspond to the shadow of the
supersymmetries of the original theory of Eq.(\ref{action1}) in $d+2$
dimensions in the presence of the (non-supersymmetric) backgrounds
$G_{MN},\Omega,W.$

Every shadow of the same theory - with the same original background
$G_{MN},\Omega,W$ taken in various gauges and parameterizations of the $d+2$
coordinates - will have the same global supersymmetry as already determined by
the SUSY condition in $d+2$ dimensions (\ref{SUSYcond}). The shadows alluded
to in this discussion are sketched in Fig.1. The same SUSY would take
different non-linear (possibly hidden) forms in terms of the coordinates in
various shadows. These shadows are all dual to each other as they retain the
information of the original theory holographically. One unchanging aspect
under the dualities is the global symmetry; in this case this includes the
SUSY determined by (\ref{SUSYcond}).

\subsubsection{Technical details \label{technical}}

In this subsection we show how the results of section \ref{shadow1} are
derived for the conformal shadow. As discussed in \cite{2tGravity}%
\cite{2tGravDetails}, we choose a convenient set of coordinates $X^{M}=\left(
w,u,x^{\mu}\right)  ,$ such that $W\left(  X\right)  =w,$ in terms of which we
will express the solutions of the kinematic equations (\ref{geom}) that
restrict the 2T geometry. See footnote (\ref{backgrounds}) for another form of
the geometry in Cartesian coordinates. It is assumed that this set of
coordinates can be chosen by coordinate reparameterizations. For example, if
the initial spacetime metric $G_{MN}$ is the flat metric $\eta_{MN}$ in $d+2 $
dimensions, the appropriate change of coordinates is given in Appendix
\ref{flat}.

We start with the solution of the kinematics for the background geometry
(\ref{geom}) as given in \cite{2tGravDetails}. The results include the
following properties of $V_{M}\equiv\frac{1}{2}\partial_{M}W$, at any $w,$
\begin{equation}%
\begin{array}
[c]{c}%
W=V^{M}V_{M}=w,\;V_{M}=\left(  \frac{1}{2},0,0\right)  _{M},\;V^{M}=\left(
2w,-\frac{1}{2},0\right)  ^{M},\;\\
V_{i}=E_{i}^{M}V_{M}=\left(  \frac{1}{2},-w,0\right)  _{i},\;V=V_{i}\Gamma
^{i}=\left(  \frac{1}{2}\Gamma^{-^{\prime}}-w\Gamma^{+^{\prime}}\right)
=\left(
\begin{array}
[c]{cc}%
0 & -i\sqrt{2}w\\
\frac{i}{\sqrt{2}} & 0
\end{array}
\right)  ,
\end{array}
\label{Vs}%
\end{equation}
The metric $G_{MN}\left(  X\right)  $ and vielbein $E_{M}^{i}\left(  X\right)
$ that satisfy (\ref{geom}) are given in terms of a general $g_{\mu\nu}\left(
x,we^{4u}\right)  $ or $e_{a}^{\text{ }\mu}\left(  x,we^{4u}\right)  $, at any
$w$, as follows
\begin{equation}
G_{MN}=%
\begin{array}
[c]{cc}%
M\backslash N &
\begin{array}
[c]{ccc}%
~~w & ~~\text{~}u~ & \text{ \ }\nu\;\;\;\;~~~~
\end{array}
\\%
\begin{array}
[c]{c}%
w\\
u\\
\mu
\end{array}
& \left(
\begin{array}
[c]{ccc}%
0 & -1 & 0\\
-1 & -4w & 0\\
0 & 0 & e^{-4u}g_{\mu\nu}%
\end{array}
\right)
\end{array}
,\;G^{MN}=%
\begin{array}
[c]{cc}%
M\backslash N &
\begin{array}
[c]{ccc}%
w & \text{~}u & \text{ \ }\nu\;\;\;\;
\end{array}
\\%
\begin{array}
[c]{c}%
w\\
u\\
\mu
\end{array}
& \left(
\begin{array}
[c]{ccc}%
4w & -1 & 0\\
-1 & 0 & 0\\
0 & 0 & e^{4u}g^{\mu\nu}%
\end{array}
\right)
\end{array}
,\label{metric}%
\end{equation}
and%
\begin{equation}
E_{M}^{\text{ \ \ }i}=%
\begin{array}
[c]{cc}%
M\backslash i &
\begin{array}
[c]{ccc}%
-^{\prime} & +^{\prime} & \text{ \ }a\;\;\;\;\;
\end{array}
\\%
\begin{array}
[c]{c}%
w\\
u\\
\mu
\end{array}
& \left(
\begin{array}
[c]{ccc}%
1 & 0 & 0\\
2w & 1 & 0\\
0 & 0 & e^{-2u}e_{\mu}^{\text{ }a}%
\end{array}
\right)
\end{array}
,\text{ }E_{i}^{\text{ }M}=%
\begin{array}
[c]{cc}%
i\backslash M &
\begin{array}
[c]{ccc}%
w & ~~\text{~}u~ & \text{ \ }\nu\;\;\;
\end{array}
\\%
\begin{array}
[c]{c}%
-^{\prime}\\
+^{\prime}\\
a
\end{array}
& \left(
\begin{array}
[c]{ccc}%
1 & 0 & 0\\
-2w & 1 & 0\\
0 & 0 & e^{2u}e_{a}^{\text{ }\mu}%
\end{array}
\right)  ,
\end{array}
\label{vielbein}%
\end{equation}
while the volume element is
\begin{equation}
d^{d+2}X\sqrt{G}\delta\left(  W\right)  =\left(  d^{d}x~du~dw\right)
e^{-2du}\sqrt{-g}\delta\left(  w\right)  .\label{vol}%
\end{equation}
The affine connection $\Gamma_{MN}^{P},$ spin connection $\omega_{M}^{ij}$ and
curvature $R_{MNP}^{Q}$ are computed in \cite{2tGravDetails}. In this paper we
will only need the expressions for $\Gamma_{wN}^{P},\Gamma_{uN}^{P}$ and
$\omega_{w}^{ij},\omega_{u}^{ij}$ $,\omega_{\mu}^{ij}$ taken from
\cite{2tGravDetails} as follows
\begin{equation}
\Gamma_{wN}^{P}=%
\begin{array}
[c]{cc}%
N\backslash P &
\begin{array}
[c]{ccc}%
\;w\; & ~\;u\;\; & \;\;\;\lambda\;\;\;\;\;\;~
\end{array}
\\%
\begin{array}
[c]{c}%
w\\
u\\
\nu
\end{array}
& \left(
\begin{array}
[c]{ccc}%
0 & 0 & 0\\
2 & \;0\; & 0\\
0 & 0 & \frac{1}{2}g^{\lambda\sigma}\partial_{w}g_{\sigma\nu}%
\end{array}
\right)
\end{array}
,\;\Gamma_{uN}^{P}=%
\begin{array}
[c]{cc}%
N\backslash P &
\begin{array}
[c]{ccc}%
\;w\; & ~\;u\;\;\;\; & \;\;\;\lambda\;\;\;\;\;\;\;\;\;\;\;\;\;\;~
\end{array}
\\%
\begin{array}
[c]{c}%
w\\
u\\
\nu
\end{array}
& \left(
\begin{array}
[c]{ccc}%
2 & 0 & 0\\
8w & -2 & 0\\
0 & 0 & -2\delta_{\nu}^{\lambda}+2wg^{\lambda\sigma}\partial_{w}g_{\sigma\nu}%
\end{array}
\right)
\end{array}
.\label{affineCo}%
\end{equation}%
\begin{equation}
\omega_{w}^{ij}=%
\begin{array}
[c]{cc}%
i\backslash j &
\begin{array}
[c]{ccc}%
-^{\prime} & +^{\prime} & \;\;b\;\;\;\;\;\;
\end{array}
\\%
\begin{array}
[c]{c}%
-^{\prime}\\
+^{\prime}\\
a
\end{array}
& \left(
\begin{array}
[c]{ccc}%
\;0\; & \;0\; & 0\\
0 & 0 & 0\\
0 & 0 & \frac{1}{2}e^{\mu\lbrack a}\partial_{w}e_{\mu}^{b]}%
\end{array}
\right)
\end{array}
,\;\;\;\omega_{u}^{ab}=%
\begin{array}
[c]{cc}%
i\backslash j &
\begin{array}
[c]{ccc}%
-^{\prime} & +^{\prime} & \;\;\;b\;\;\;\;\;\;\;\;\;
\end{array}
\\%
\begin{array}
[c]{c}%
-^{\prime}\\
+^{\prime}\\
a
\end{array}
& \left(
\begin{array}
[c]{ccc}%
\;0\; & -2\; & 0\\
2 & 0 & 0\\
0 & 0 & 2we^{\mu\lbrack a}\partial_{w}e_{\mu}^{b]}%
\end{array}
\right)
\end{array}
.\label{spinCo}%
\end{equation}
and
\begin{equation}
\omega_{\mu}^{ij}=%
\begin{array}
[c]{cc}%
i\backslash j &
\begin{array}
[c]{ccc}%
\;\;\;\text{\ \ \ \ }-^{\prime}\;\;\;\; & \;\;\;\;\;\;\;\;\;\;\;\;\text{
\ }+^{\prime}\text{ \ \ \ \ \ \ \ \ \ \ \ \ \ \ } & \text{ \ \ \ \ }%
\;b\;\;\;\;\;\;\;\;\;\;\;\;\;\;\;\;\;\;\;
\end{array}
\\%
\begin{array}
[c]{c}%
-^{\prime}\\
+^{\prime}\\
a
\end{array}
& \left(
\begin{array}
[c]{ccc}%
0 & 0 & e^{-2u}\left(  -2e_{\mu}^{b}+we^{~b\sigma}\partial_{w}g_{\mu\sigma
}\right) \\
0 & 0 & \frac{e^{-2u}}{2}e^{~b\nu}\partial_{w}g_{\mu\nu}\\
e^{-2u}\left(  2e_{\mu}^{~a}-we^{~a\sigma}\partial_{w}g_{\lambda\sigma}\right)
& -\frac{e^{-2u}}{2}e^{~a\sigma}\partial_{w}g_{\mu\sigma} & \omega_{\mu}%
^{ab}\left(  e\right)
\end{array}
\right)
\end{array}
\label{spinComu}%
\end{equation}
where $\omega_{\mu}^{ab}\left(  e\right)  $ is the standard spin connection
constructed from the vielbein $e_{\mu}^{a}$ in $d$ dimensions$.$ It is
interesting that all dependence on $x^{\mu}$ and $we^{4u}$ drops out in the
following combination of connections
\begin{equation}
V^{M}\Gamma_{MN}^{P}=2w\Gamma_{wN}^{P}-\frac{1}{2}\Gamma_{uN}^{P}=\delta
_{M}^{P}-2\delta_{w}^{P}\delta_{M}^{w},\;\;\;V^{M}\omega_{M}^{ij}=2w\omega
_{w}^{ij}-\frac{1}{2}\omega_{u}^{ij}=-\delta_{+^{\prime}}^{[i}\delta
_{-^{\prime}}^{j]}.\label{simpConn}%
\end{equation}

In this basis the kinematic equations for the dilaton and the SUSY parameter
Eqs.(\ref{geom},\ref{kineps}) simplify to $\left(  2w\partial_{w}-\frac{1}%
{2}\partial_{u}+\frac{d-2}{2}\right)  \Omega\left(  X\right)  =0$ and $\left(
2wD_{w}-\frac{1}{2}D_{u}\right)  \varepsilon\left(  X\right)  =\left(
2w\partial_{w}-\frac{1}{2}\partial_{u}+\frac{1}{2}\Gamma^{+^{\prime}-^{\prime
}}\right)  \varepsilon\left(  X\right)  =0$ respectively. These restrict the
$u,w$ dependence of the dilaton and the SUSY parameter as follows
\begin{equation}
\Omega\left(  X\right)  =e^{\left(  d-2\right)  u}\hat{\Omega}\left(
x,we^{4u}\right)  ,\;\;\varepsilon\left(  X\right)  =\exp\left(
u\Gamma^{+^{\prime}-^{\prime}}\right)  \hat{\varepsilon}\left(  x,we^{4u}%
\right)  .\label{omeps}%
\end{equation}
The backgrounds $G_{MN},E_{M}^{i},\omega_{M}^{ij},\Gamma_{MN}^{P},\Omega$
occur in our action without further derivatives with respect to $u$ or $w,$
while there is a delta function $\delta\left(  w\right)  $ in the volume
element (\ref{vol}). So, in (\ref{metric}-\ref{omeps}), considering a Taylor
expansion in powers of $we^{4u},$ we must keep only the zeroth order terms
since all higher order terms in $w$ drop out due to $w^{p}\delta\left(
w\right)  =0$ for integers $p\geq1.$

Now we turn to the solution of the kinematic constraints (\ref{kin}) for the
dynamical fields $A_{M}^{a},\lambda_{A}^{a}$. We will follow a procedure
similar to sections 4B and 4C in \cite{2tstandardM} except for generalizing to
curved space and higher dimensions. We will work in the Yang-Mills gauge given
by $V\cdot A^{a}=2wA_{w}^{a}-\frac{1}{2}A_{u}^{a}=0.$ In this gauge there
remains a subset of Yang-Mills gauge symmetry which does not change the gauge
$V\cdot A^{a}=0.$ For this subset the gauge parameters $\Lambda^{a}\left(
X\right)  $ satisfy $0=V\cdot\delta_{\Lambda}A^{a}=V\cdot D\Lambda^{a}%
=V\cdot\partial\Lambda^{a}=\left(  2w\partial_{w}-\frac{1}{2}\partial
_{u}\right)  \Lambda^{a}.$ So the remaining Yang-Mills gauge symmetry has the
form
\begin{equation}
\Lambda^{a}\left(  X\right)  =\hat{\Lambda}^{a}\left(  x,we^{4u}\right)
,\label{remainYM}%
\end{equation}
with $\hat{\Lambda}^{a}$ an arbitrary function of $x$ and $we^{4u}.$ This will
be used for further Yang-Mills gauge fixing.

In the gauge $V\cdot A^{a}=0$ the interaction terms with the Yang-Mills field
disappear in the kinematic constraints (\ref{kin}) $0=V^{M}F_{MN}^{a}=\left(
V\cdot\nabla+1\right)  A_{N}^{a}$, where we have also used $\nabla_{M}%
V_{N}=G_{MN}$ (see \ref{geom}) to pass $V$ through $\partial$. So, we have
\begin{equation}
0=\left(  V\cdot\nabla+1\right)  A_{N}^{a}=\left(  2w\partial_{w}-\frac{1}%
{2}\partial_{u}+1\right)  A_{N}^{a}-\left(  2w\Gamma_{wN}^{P}-\frac{1}%
{2}\Gamma_{uN}^{P}\right)  A_{P}^{a}\;.
\end{equation}
Taking into account Eq.(\ref{simpConn}), the solution of this kinematic
equation for $A_{M}^{a}\left(  X\right)  $ is
\begin{equation}
A_{w}^{a}\left(  X\right)  =e^{4u}\hat{A}_{w}^{a}\left(  x,we^{4u}\right)
,\;A_{u}^{a}\left(  X\right)  =\hat{A}_{u}^{a}\left(  x,we^{4u}\right)
,\;A_{\mu}^{a}\left(  X\right)  =\hat{A}_{\mu}^{a}\left(  x,we^{4u}\right)
.\label{Asolved}%
\end{equation}
Similarly, the kinematic constraint for the spinor is solved as follows
(taking into account Eq.(\ref{simpConn}))
\begin{align}
0 &  =\left(  V\cdot D+\frac{d}{2}\right)  \lambda^{a}=\left(  2wD_{w}%
-\frac{1}{2}D_{u}+\frac{d}{2}\right)  \lambda^{a}\\
&  \Rightarrow\;\lambda^{a}=\exp\left(  ud+u\Gamma^{+^{\prime}-^{\prime}%
}\right)  \hat{\lambda}^{a}\left(  x,we^{4u}\right)  .\label{Lsolved}%
\end{align}
Next we recall that the action (\ref{action1}) and the SUSY current
(\ref{current}) are gauge invariant under the 2T gauge
transformations\footnote{The gauge symmetry of the action with fields
completely off shell involves more complicated transformation rules than the
one shown here, and in that case the parameters $s_{M}^{a}\left(  X\right)  $,
$\kappa_{1A}^{a}\left(  X\right)  $, $\kappa_{2A}^{a}\left(  X\right)  $ are
arbitrary. However, in the present case the fields $A_{M}^{a}\left(  X\right)
,\lambda_{A}^{a}\left(  X\right)  $ in (\ref{Asolved},\ref{Lsolved}) already
satisfy the kinematic constraints (\ref{kin}), so the corresponding local
parameters $s_{M}^{a}\left(  X\right)  $, $\kappa_{1A}^{a}\left(  X\right)  $,
$\kappa_{2A}^{a}\left(  X\right)  $ must be specialized to a subset that is
consistent with the kinematic constraints. Hence we must have $s_{w}%
^{a}\left(  X\right)  =\hat{s}_{w}^{a}\left(  x,we^{4u}\right)  ,\;s_{u}%
^{a}\left(  X\right)  =e^{4u}\hat{s}_{u}^{a}\left(  x,we^{4u}\right)  $ and
$s_{\mu}^{a}\left(  X\right)  =e^{4u}\hat{s}_{\mu}^{a}\left(  x,we^{4u}%
\right)  $ and similarly $\kappa_{1}^{a}\left(  X\right)  =\exp\left(
u\Gamma^{+^{\prime}-^{\prime}}+u\left(  d+2\right)  \right)  \hat{\kappa}%
_{1}^{a}\left(  x,we^{4u}\right)  $, and $\kappa_{2}^{a}\left(  X\right)
=\exp\left(  u\Gamma^{+^{\prime}-^{\prime}}+u\left(  d+4\right)  \right)
\hat{\kappa}_{2}^{a}\left(  x,we^{4u}\right)  .$ Under such specialized
transformations, as in the more general case, the kinematic constraints
(\ref{kin}) on $A_{M}^{a}\left(  X\right)  ,\lambda_{A}^{a}\left(  X\right)  $
as well as the action or the dynamical equations (\ref{dyn}) are invariant.
\label{specialized}} $\delta_{s}A_{M}^{a}=Ws_{M}^{a}\left(  X\right)  $ and
$\delta_{\kappa}\lambda^{a}=V\kappa_{1}^{a}\left(  X\right)  +W\kappa_{2}%
^{a}\left(  X\right)  ,$ with local bosonic parameters $s_{M}^{a}\left(
X\right)  $ and fermionic parameters $\kappa_{1A}^{a}\left(  X\right)
,\kappa_{2A}^{a}\left(  X\right)  .$ From this we deduce that the
kinematically constrained fields above transform under these gauge symmetries
as follows
\begin{equation}%
\begin{array}
[c]{c}%
\delta_{s}\hat{A}_{M}^{a}=we^{4u}\hat{s}_{M}^{a}\left(  x,we^{4u}\right)
,\;\;\\
\delta_{\kappa}\hat{\lambda}^{a}=\left(  \frac{1}{2}\Gamma^{-^{\prime}%
}-we^{4u}\Gamma^{+^{\prime}}\right)  \hat{\kappa}_{1}^{a}\left(
x,we^{4u}\right)  +we^{4u}\hat{\kappa}_{2}^{a}\left(  x,we^{4u}\right)  .
\end{array}
\end{equation}
This is enough gauge symmetry at any $w$ to gauge fix $\hat{A}_{M}^{a}\left(
x,we^{4u}\right)  ,\hat{\lambda}_{A}^{a}\left(  x,we^{4u}\right)  $ to
functions of only $x$ (it may be helpful for the reader to contemplate an
expansion in powers of $we^{4u}$)$,$ while eliminating half of the spinor
degrees of freedom with the gauge choice\footnote{Here we use explicit gamma
matrices that satisfy $\Gamma^{i}\bar{\Gamma}^{j}+\Gamma^{j}\bar{\Gamma}%
^{i}=2\eta^{ij}:$ $\Gamma^{+^{\prime}}=\left(
%TCIMACRO{\QATOP{0}{0}}%
%BeginExpansion
\genfrac{}{}{0pt}{}{0}{0}%
%EndExpansion%
%TCIMACRO{\QATOP{-i\sqrt{2}}{0}}%
%BeginExpansion
\genfrac{}{}{0pt}{}{-i\sqrt{2}}{0}%
%EndExpansion
\right)  $, $\Gamma^{-^{\prime}}=\left(
%TCIMACRO{\QATOP{0}{-i\sqrt{2}}}%
%BeginExpansion
\genfrac{}{}{0pt}{}{0}{-i\sqrt{2}}%
%EndExpansion%
%TCIMACRO{\QATOP{0}{0}}%
%BeginExpansion
\genfrac{}{}{0pt}{}{0}{0}%
%EndExpansion
\right)  $, $\Gamma^{\mu}=\left(
%TCIMACRO{\QATOP{\bar{\gamma}^{\mu}}{0}}%
%BeginExpansion
\genfrac{}{}{0pt}{}{\bar{\gamma}^{\mu}}{0}%
%EndExpansion%
%TCIMACRO{\QATOP{0}{-\gamma^{\mu}}}%
%BeginExpansion
\genfrac{}{}{0pt}{}{0}{-\gamma^{\mu}}%
%EndExpansion
\right)  $ with\ $\gamma^{\mu}=\left(  -1,\gamma^{i}\right)  $,\ \ $\bar
{\gamma}^{\mu}=\left(  1,\gamma^{i}\right)  $ and $\bar{\Gamma}^{+^{\prime}%
}=\left(
%TCIMACRO{\QATOP{0}{0}}%
%BeginExpansion
\genfrac{}{}{0pt}{}{0}{0}%
%EndExpansion%
%TCIMACRO{\QATOP{-i\sqrt{2}}{0}}%
%BeginExpansion
\genfrac{}{}{0pt}{}{-i\sqrt{2}}{0}%
%EndExpansion
\right)  $,\ \ $\bar{\Gamma}^{-^{\prime}}=\left(
%TCIMACRO{\QATOP{0}{-i\sqrt{2}}}%
%BeginExpansion
\genfrac{}{}{0pt}{}{0}{-i\sqrt{2}}%
%EndExpansion%
%TCIMACRO{\QATOP{0}{0}}%
%BeginExpansion
\genfrac{}{}{0pt}{}{0}{0}%
%EndExpansion
\right)  $,\ $\bar{\Gamma}^{\mu}=\left(
%TCIMACRO{\QATOP{\gamma^{\mu}}{0}}%
%BeginExpansion
\genfrac{}{}{0pt}{}{\gamma^{\mu}}{0}%
%EndExpansion%
%TCIMACRO{\QATOP{0}{-\bar{\gamma}^{\mu}}}%
%BeginExpansion
\genfrac{}{}{0pt}{}{0}{-\bar{\gamma}^{\mu}}%
%EndExpansion
\right)  .$ In this basis the conjugate of $\Psi=\left(
%TCIMACRO{\QATOP{\psi_{1}}{\psi_{2}}}%
%BeginExpansion
\genfrac{}{}{0pt}{}{\psi_{1}}{\psi_{2}}%
%EndExpansion
\right)  $ is given by $\bar{\Psi}=i\left(  \bar{\psi}_{2},\;\bar{\psi}%
_{1}\right)  .$ See appendix A in \cite{susy2tN1}. \label{gammas}}
$\Gamma^{+^{\prime}}\hat{\lambda}^{a}=0$
\begin{equation}
\hat{A}_{M}^{a}\left(  x,we^{4u}\right)  \rightarrow A_{M}\left(  x\right)
,\;\;\hat{\lambda}_{A}^{a}\left(  x,we^{4u}\right)  \rightarrow\left(
\begin{array}
[c]{c}%
\lambda_{\alpha}\left(  x\right) \\
0
\end{array}
\right)  ,\;\overline{\hat{\lambda}^{a}}\left(  x,we^{4u}\right)  \rightarrow
i\left(  0\;\;\overline{\lambda^{a}}\left(  x\right)  \right)  .
\end{equation}
Now, recall that $V\cdot A=0$ implied that $A_{u}(X)=2wA_{w}\left(  X\right)
. $ Inserting the forms in (\ref{Asolved}) this gives $\hat{A}_{u}%
=2we^{4u}\hat{A}_{w}\left(  x,we^{4u}\right)  .$ We can now use the remaining
Yang-Mills symmetry in (\ref{remainYM}) to fix further the gauge $\hat{A}%
_{w}\left(  x,we^{4u}\right)  =0,$ which then also makes $\hat{A}_{u}=0.$

All of the above steps solve the kinematic constraints at any $w.$ Having
taken into account all derivatives with respect to $w,$ now we can safely set
$w=0$ on account that the volume element in the action contains the delta
function $\delta\left(  w\right)  $ as in (\ref{vol}). So, the above fields
now are taken at $w=0,$ yielding only functions of spacetime $x^{\mu} $ in $d$
dimensions. The $u$ dependence of all fields is explicit as in
(\ref{fieldShadows}), and after inserting them into the action one finds that
all $u$ dependence of the Lagrangian $L_{SYM}$ cancels out against the $u$
dependence of the volume element (\ref{vol}), leaving an action density that
is independent of $u$. Then the integral over $u$ in the action is an overall
infinite factor which is absorbed into the overall renormalization constant
$K$ in front of the action (\ref{action1}). Equivalently, this is a
renormalization of the Planck constant in the path integral formalism.

In summary, by a series of gauge choices and solving kinematic constraints we
end up with the configurations in Eq.(\ref{fieldShadows}) which describe the
shadow fields in $d$ dimensions. Inserting these in the original action
(\ref{action1}) we obtain the results summarized in the shadow action
(\ref{actionShadow}) and the comments that follow it.

\subsection{Dimensional reduction SYM$_{10+2}^{1}$ $\rightarrow$
SYM$_{4+2}^{4}$ $\rightarrow$ shadow SYM$_{3+1}^{4}$\label{gen4+2}}

Let us now consider the reduction 10+2 $\rightarrow4+2$ by taking the fields
as a functions of the coordinates%
\begin{equation}
X^{M}=\left(  x^{m},y^{I}\right)  ,\;\left\{
\begin{array}
[c]{l}%
x^{m}\text{ vector of SO}\left(  4,2\right)  ,\\
y^{I}\text{ vector of SO}\left(  6\right)  .
\end{array}
\right.
\end{equation}
We are aiming for a metric $G_{MN}\left(  x,y\right)  $ consistent with
SO$\left(  4,2\right)  \times$SO$\left(  6\right)  $ symmetry, but the overall
metric need not be flat in $10+2$ dimensions. In fact we will see that to
recover SYM$_{4+2}^{1}$ via compactification, with 32 supersymmetries, the
metric $G_{MN}\left(  x,y\right)  $ cannot be flat.

\subsubsection{Background consistent with homothety and SUSY}

We take a metric and vielbein of the form
\begin{equation}
G_{MN}=%
\begin{array}
[c]{c}%
m\\
I
\end{array}
\overset{%
\begin{array}
[c]{cc}%
n\;\;\; & \;\;J
\end{array}
}{\left(
\begin{array}
[c]{cc}%
\;\;\eta_{mn}\;\; & 0\\
0 & a^{2}\left(  x,y\right)  ~\delta_{IJ}%
\end{array}
\right)  },\;E_{M}^{i}=%
\begin{array}
[c]{c}%
m\\
I
\end{array}
\overset{%
\begin{array}
[c]{cc}%
\alpha\;\;\; & \;\;a
\end{array}
}{\left(
\begin{array}
[c]{cc}%
\;\;\delta_{m}^{~\alpha}\;\; & 0\\
0 & a\left(  x,y\right)  ~\delta_{I}^{a}%
\end{array}
\right)  }.\label{metric66}%
\end{equation}
which is flat in $4+2$ dimensions $x^{m}$, and conformally flat in the extra 6
dimensions $y^{I}$ due to the warp factor $a^{2}\left(  x,y\right)  $.
Furthermore, by choosing coordinates such that $W\left(  X\right)  =x^{2}$ we
get%
\begin{equation}
V^{M}\left(  x,y\right)  =\frac{1}{2}G^{MN}\partial_{M}W=\overset
{m\;\;\;\;I}{\left(  x^{m},\;0\right)  }\text{ and~ }V^{i}=V^{M}E_{M}%
^{i}=\overset{\alpha\;\;\;\;\;\;\;\;a}{\left(  \delta_{m}^{\alpha}%
x^{m},\;0\right)  }.
\end{equation}
Next impose the homothety conditions for the metric and vielbein (\ref{geom})
\begin{equation}
\pounds _{V}G_{MN}=2G_{MN},\;\pounds _{V}E_{M}^{i}=E_{M}^{i}.
\end{equation}
Specializing $M\rightarrow m,I$ we learn that the conformal factor $a\left(
x,y\right)  $ must be homogeneous and satisfies the equation
\begin{equation}
x\cdot\partial\ln a=1\text{, or }a\left(  tx,y\right)  =ta\left(  x,y\right)
.\label{lna}%
\end{equation}
Similarly, the homothety condition (\ref{geom}) for the dilaton $\Omega$
reduces to $\left(  x\cdot\partial+\frac{d-2}{2}\right)  \Omega=0,$ which
requires a homogeneous dilaton
\begin{equation}
\Omega\left(  tx,y\right)  =t^{-4}\Omega\left(  x,y\right)  \text{ for
}d+2=12.
\end{equation}
The spin connection $\omega_{M}^{ij}$ has to reproduce the $E_{M}^{i}$ above
through $E_{M}^{i}=D_{M}V^{i}$. Hence the spin connection can be taken as%
\begin{equation}
\omega_{M}^{ij}\left(  x,y\right)  =%
\begin{tabular}
[c]{|c|c|c|c|}\hline
$_{M}$
%TCIMACRO{\TEXTsymbol{\backslash}}%
%BeginExpansion
$\backslash$%
%EndExpansion
$~~ij$ & $^{\alpha\beta}$ & $^{\beta a}$ & $^{ab}$\\\hline
$\omega_{m}^{ij}=$ & $\;\;\;\;\;0\;\;\;\;\;$ & $0$ & $0$\\\hline
$\omega_{I}^{ij}=$ & $0$ & $\omega_{I}^{\beta a}=\delta_{I}^{a}\delta^{\beta
m}\partial_{m}\ln a$ & $~~~\omega_{I}^{ab}=\delta_{I}^{[a}\delta^{b]J}\left(
\partial_{J}\ln a\right)  $\\\hline
\end{tabular}
\label{spinCon}%
\end{equation}
With these $\omega_{M}^{ij}\left(  x,y\right)  $ the torsion tensor vanishes,
as it should, $T_{MN}^{i}=D_{[M}E_{N]}^{i}=\partial_{\lbrack M}E_{N]}%
^{i}+\omega_{\lbrack M}^{ij}E_{N]j}=R_{MN}^{ij}V_{j}=0.$

\subsubsection{Reduction of the 10+2 action to 4+2}

Now consider the action. We are aiming to obtain the dimensionally reduced
action to coincide with SYM$_{4+2}^{4}$ whose action was given in
\cite{susy2tN2N4}. The Yang-Mills field $A_{M}(X)=\left(  A_{m},A_{I}\right)
\left(  x\right)  $ is taken independent of $y^{I}$ due to the dimensional
reduction. Hence in constructing $F_{MN}=\left(  F_{mn},F_{mI},F_{IJ}\right)
$ all derivatives with respect to $y^{I}$ are dropped, so that
\begin{equation}
F_{mI}=D_{m}A_{I}=\partial_{m}A_{I}+A_{m}\times A_{I};\;\;F_{IJ}=A_{I}\times
A_{J},
\end{equation}
where $A_{m}\times A_{I}$ is a short hand notation for the adjoint action of
the Yang-Mills group,
\begin{equation}
\left(  A_{m}\times A_{I}\right)  _{a}\equiv f_{abc}A_{m}^{b}A_{I}%
^{c},\;\;\text{etc.}%
\end{equation}
Then, the Yang-Mills term in Eq.(\ref{action}) becomes%
\begin{align}
&  -\frac{1}{4g_{YM}^{2}}\delta\left(  W\right)  \Omega^{3/2}\sqrt{G}%
G^{MP}G^{NQ}F_{MN}F_{NQ}\nonumber\\
&  =-\frac{1}{4g_{YM}^{2}}\delta\left(  x^{2}\right)  \Omega^{3/2}a^{6}\left(
\left(  F_{mn}\right)  ^{2}+2a^{-2}\left(  D_{m}A_{I}\right)  ^{2}%
+a^{-4}\left(  A_{I}\times A_{J}\right)  ^{2}\right) \nonumber\\
&  =-\frac{1}{4g_{YM}^{2}}\delta\left(  x^{2}\right)  \Omega^{3/2}a^{6}\left(
\left(  F_{mn}\right)  ^{2}+2\left(  D_{m}\frac{A_{I}}{a}+\frac{A_{I}}%
{a}\partial_{m}\ln a\right)  ^{2}+\left(  \frac{A_{I}}{a}\times\frac{A_{J}}%
{a}\right)  ^{2}\right) \label{YM}%
\end{align}
Here we will identify $A_{I}/a$ with the six scalar fields in SYM$_{4+2}^{4}$%
\begin{equation}
\phi_{I}\left(  x\right)  =\frac{1}{g_{YM}}\frac{A_{I}\left(  x\right)
}{a\left(  x,y\right)  },\text{ vector of SO}\left(  6\right)  .
\end{equation}
Since $\phi_{I}\left(  x\right)  $ must be independent of $y^{I}$ we must take
the warp factor $a\left(  x,y\right)  $ independent of $y^{I}.$ The kinetic
term for $\phi_{I}\left(  x\right)  $ coming from the reduction from 10+2
contains the form%
\begin{equation}
D_{m}\left(  \frac{A_{I}}{ag_{YM}}\right)  +\left(  \frac{A_{I}}{ag_{YM}%
}\right)  \partial_{m}\ln a=\left(  D_{m}\phi_{I}+\phi_{I}\partial_{m}\ln
a\right)  .
\end{equation}
So the kinetic term for $\phi_{I}$ in Eq.(\ref{YM}) becomes (for all
contractions over $m$ we use the flat SO$\left(  4,2\right)  $ metric)
\begin{align}
&  -\frac{1}{2}\delta\left(  x^{2}\right)  \left(  D_{m}\phi_{I}+\phi
_{I}\partial_{m}\ln a\right)  ^{2}\nonumber\\
&  =-\frac{1}{2}\delta\left(  x^{2}\right)  \left(  \left(  D_{m}\phi
_{I}\right)  ^{2}+\partial_{m}\ln a\partial_{m}\phi^{2}+\phi^{2}\left(
\partial_{m}\ln a\right)  ^{2}\right) \nonumber\\
&  =\left\{
\begin{array}
[c]{c}%
+\frac{1}{2}\delta\left(  x^{2}\right)  \phi_{I}D_{m}^{2}\phi_{I}+\frac{1}%
{2}\phi^{2}\left\{
\begin{array}
[c]{c}%
\delta^{\prime}\left(  x^{2}\right)  \left(  -2+2x\cdot\partial\ln a\right) \\
+\delta\left(  x^{2}\right)  \left(  \partial_{m}^{2}\ln a-\left(
\partial_{m}\ln a\right)  ^{2}\right)
\end{array}
\right\} \\
+\frac{1}{2}\partial^{m}\left[  -\delta\left(  x^{2}\right)  \phi_{I}D_{m}%
\phi_{I}+\delta^{\prime}\left(  x^{2}\right)  x_{m}\phi^{2}-\delta\left(
x^{2}\right)  \left(  \partial_{m}\ln a\right)  \phi^{2}\right]
\end{array}
\right\} \label{totalDiv}%
\end{align}
Here the last term is a total derivative and can be dropped in the action. To
obtain this form we used $\partial_{m}x^{m}=6$ and $x\cdot\partial
\delta^{\prime}\left(  x^{2}\right)  =-4\delta^{\prime}\left(  x^{2}\right)
.$ Using $x\cdot\partial\ln a=1$ in Eq.(\ref{lna}) the coefficient of
$\delta^{\prime}\left(  x^{2}\right)  $ vanishes, $\left(  -2+2x\cdot
\partial\ln a\right)  =0$. Hence the kinetic term for the scalar field is%
\begin{equation}
\delta\left(  x^{2}\right)  \Omega^{3/2}a^{6}\left\{  \frac{1}{2}\phi_{I}%
D^{2}\phi_{I}+\frac{1}{2}\phi_{I}^{2}\left(  \partial_{m}^{2}\ln a-\left(
\partial_{m}\ln a\right)  ^{2}\right)  \right\}  .\label{kinScalar}%
\end{equation}
The last term could be interpreted as a coupling to a background curvature in
4+2 dimensions, but we will continue here under the assumption that the 4+2
background is flat since we are trying to compare to the SYM$_{4+2}^{4}$ in
\cite{susy2tN2N4}. Hence we need to impose $\partial_{m}^{2}\ln a-\left(
\partial_{m}\ln a\right)  ^{2}=0.$ So the coefficient of $\phi^{2}%
\delta\left(  x^{2}\right)  $ vanishes only when $a\left(  x\right)  $
satisfies the following solution%
\begin{equation}
\partial^{2}\ln a-\left(  \partial_{m}\ln a\right)  ^{2}=0~~\rightarrow\text{
~}a\left(  x\right)  =x\cdot b\text{ and }b^{m}b_{m}=0.
\end{equation}
To get the normalizations of the first and second terms in Eq.(\ref{YM}) to
coincide with \cite{susy2tN2N4} we must also have $\Omega^{3/2}a^{6}=1.$ Hence
$a\left(  x,y\right)  ,\Omega\left(  x,y\right)  $ should both be independent
of $y^{I}$ and related to each other as
\begin{equation}
a\left(  x\right)  =\Omega^{-\frac{1}{4}}\left(  x\right)  =x\cdot
b.\label{aOmeg}%
\end{equation}
A tricky term in the SYM$_{4+2}^{4}$ action is the kinetic term for the scalar
that has the form
\begin{equation}
\frac{1}{2}\delta\left(  x^{2}\right)  \phi_{I}D_{m}^{2}\phi_{I}%
\end{equation}
rather than $-\frac{1}{2}\delta\left(  x^{2}\right)  \eta^{mn}D_{m}\phi
_{I}D_{n}\phi_{I}.$ These are not the same because an integration by parts
involves a difference term proportional to $\delta^{\prime}\left(
x^{2}\right)  .$ This form of the kinetic term for scalars is required by both
the 2T gauge symmetries and the SUSY symmetry in $4+2$ dimensions (for the
most general form permitted in the presence of curved backgrounds see
\cite{2tScalars}). Then, with the form of $a\left(  x\right)  $ in
Eq.(\ref{aOmeg}), we obtain the correct kinetic term for the scalars%
\begin{equation}
-\frac{1}{2}\delta\left(  x^{2}\right)  \left(  D_{m}\phi_{I}+\phi_{I}%
\partial_{m}\ln a\right)  ^{2}=\frac{1}{2}\delta\left(  x^{2}\right)  \phi
_{I}D^{2}\phi_{I}+\text{total derivative.}%
\end{equation}
Note that the constant vector $b_{m}$ has disappeared from all terms. So there
is no preferred direction in the resulting action and therefore there is an
SO$\left(  4,2\right)  \times$SO$\left(  6\right)  $ symmetry.

Next we consider the fermions. For correct normalization, the fermion must be
taken as%
\begin{equation}
\lambda_{A}\left(  x,y\right)  =a^{-3}\left(  x\right)  \psi_{A}\left(
x\right)  ,\;A=1,2,\cdots,32.
\end{equation}
Then the fermion action becomes%
\begin{align}
&  \frac{i}{2}\delta\left(  W\right)  \sqrt{G}\bar{\lambda}VD\lambda+h.c\\
&  =\frac{i}{2}\delta\left(  x^{2}\right)  \sqrt{G}a^{-3}\bar{\psi}x\left(
\Gamma^{m}D_{m}+\Gamma^{a}E_{a}^{I}\left(  \partial_{I}+\omega_{I}+A_{I}%
\times\right)  \right)  \left(  \psi a^{-3}\right)  +h.c\\
&  =\frac{i}{2}\delta\left(  x^{2}\right)  \left(  a^{6}a^{-6}\right)
\bar{\psi}x\left(
\begin{array}
[c]{c}%
\Gamma^{m}D_{m}+\Gamma^{m}\partial_{m}\ln a^{-3}\\
+\frac{1}{a}\Gamma^{I}\left(
\begin{array}
[c]{c}%
\partial_{I}\ln a^{-3}-\frac{1}{2}\Gamma_{\beta}\Gamma_{a}\omega_{I}^{\beta
a}\\
+\frac{1}{4}\Gamma_{cd}\omega_{I}^{cd}+A_{I}\times
\end{array}
\right)
\end{array}
\right)  \psi\left(  x\right)  +h.c
\end{align}
After taking into account that $a$ is independent of $y^{I}$ we can drop
$\omega_{I}^{cd}=\delta_{I}^{[c}\delta^{d]J}\left(  \partial_{J}\ln a\right)
=0$ and write $\omega_{I}^{\beta a}=\delta_{I}^{a}\delta^{\beta m}\partial
_{m}\ln a,\;$we note that
\begin{equation}
-3\Gamma^{m}\partial_{m}\ln a+\frac{1}{2}\Gamma^{I}\Gamma^{I}\Gamma
^{m}\partial_{m}\ln a=0.\label{cancelletion}%
\end{equation}
Hence we get the correct kinetic term for fermions that agrees with the
expected form for SYM$_{4+2}^{4}$ in agreement with \cite{susy2tN2N4}%
\[
\frac{i}{2}\delta\left(  W\right)  \sqrt{G}\bar{\lambda}VD\lambda=\frac{i}%
{2}\delta\left(  x^{2}\right)  \bar{\psi}\left[  x\left(  \Gamma^{m}%
D_{m}+g\Gamma^{I}\phi_{I}\times\right)  \right]  \psi\left(  x\right)
\]

Putting together the result of the reduction, and dropping the total
derivative in Eq.(\ref{totalDiv}), we obtain the reduced Lagrangian
\begin{equation}
L_{SYM}\left(  x,y\right)  =\delta\left(  x^{2}\right)  \left(
\begin{array}
[c]{c}%
-\frac{1}{4g_{YM}^{2}}\left(  F_{mn}\right)  ^{2}+\frac{1}{2}\phi_{I}D^{2}%
\phi_{I}-\frac{g_{YM}^{2}}{4}\left(  \phi_{I}\times\phi_{J}\right)  ^{2}\\
+\frac{i}{2}\bar{\psi}\left[  x\left(  \bar{\Gamma}^{m}D_{m}+g_{YM}\bar
{\Gamma}^{I}\phi_{I}\times\right)  \right]  \psi\left(  x\right)  +h.c
\end{array}
\right) \label{reduced42}%
\end{equation}
The $y$ integration over a compact space is an overall trivial factor that can
be absorbed into the normalization $K$ in the original action (\ref{action1}).
We know from \cite{susy2tN2N4} that SYM$_{4+2}^{4}$ is the 2T-physics parent
of SYM$_{3+1}^{4}$, hence we have established the connections shown with
arrows on the right hand side of Fig.1.

In this last form the fermions are still retaining the 10+2 notation for the
32 $\lambda^{\prime}s$ as the spinor of SO$\left(  10,2\right)  ,$ while the
gamma matrices $\Gamma^{m},\Gamma^{I}$ are also 32$\times32$ matrices, thus
\textit{showing their 10+2 dimensional origin}. To relate to the spinors in
$4+2$ dimensions and to display the $\mathcal{N}=4$ supersymmetry we must
express the 32-spinor in an SU$\left(  2,2\right)  \times$SU$\left(  4\right)
$ basis as in \cite{susy2tN2N4}. This is a technical point in group theory but
may be useful to show it explicitly as in Appendix (\ref{32s}).

Using the notation of 32$\times32$ gamma matrices provided in Appendix
(\ref{32s}), it is straightforward to show that the fermion kinetic term in
the SO$\left(  10,2\right)  $ notation with the 32 $\lambda^{\prime}s$ is
rewritten correctly in the SO$\left(  4,2\right)  \times$SO$\left(  6\right)
=$SU$\left(  2,2\right)  \times$SU$\left(  4\right)  $ basis in agreement with
Eq.(4.1) in \cite{susy2tN2N4} (the Yang-Mills group adjoint label $a$ is now
shown explicitly below, while the label $r$ is for the $4$ of SU$\left(
4\right)  $ as defined in the appendix)
\begin{equation}
L^{\mathcal{N}=4}\left(  x\right)  =\delta\left(  x^{2}\right)  \left\{
\begin{array}
[c]{c}%
-\frac{1}{4g_{YM}^{2}}F_{mn}^{a}F_{a}^{mn}+\frac{1}{2}\phi_{I}^{a}D^{m}%
D_{m}\phi_{I}^{a}-\frac{g_{YM}^{2}}{4}%
%TCIMACRO{\dsum }%
%BeginExpansion
{\displaystyle\sum}
%EndExpansion
\left(  f_{abc}\phi_{I}^{b}\phi_{J}^{c}\right)  ^{2}\\
+\frac{i}{2}\left[  \bar{\psi}^{ar}x\bar{D}\psi_{r}^{a}+g_{YM}f_{abc}\left(
\psi_{r}^{a}C\bar{x}\psi_{s}^{b}\right)  \left(  \bar{\gamma}^{I}\right)
^{rs}\phi_{I}^{c}\right]  +h.c.
\end{array}
\right\} \label{n4action1}%
\end{equation}
In \cite{susy2tN2N4} it is shown how to rewrite the kinetic and potential
energies of the six real scalars $\phi_{I}^{a}$ in an SU$\left(  4\right)  $
antisymmetric pseudo-complex matrix notation $\varphi_{rs}=\left(  \gamma
^{I}\right)  _{rs}\phi_{I}^{c}=\frac{1}{2}\varepsilon_{rstu}\bar{\varphi}%
^{tu}.$ This SU$\left(  4\right)  $ notation displays the \textit{linearly
realized }SU$\left(  2,2|4\right)  $\textit{\ supersymmetry} of the
SYM$_{4+2}^{4}$ theory directly in $4+2$ dimensions. This is the origin of the
superconformal symmetry that is \textit{non-linearly} realized in the
conformal shadow in the form of the conventional SYM$_{3+1}^{4}$ theory shown
at the bottom of Fig.1.

\section{M(atrix) theory as dimensionally reduced SYM$_{10+2}^{1}$}

It is well known that M(atrix) theory in 9+1 dimensions is constructed by
compactifying SYM$_{9+1}^{1}$ \cite{matrixTh1}-\cite{matrixThJ4}. Since we
have already shown in section (\ref{shadow1}) that SYM$_{9+1}^{1}$ is a shadow
of SYM$_{10+2}^{1},$ it is already evident that SYM$_{10+2}^{1}$ is the
2T-physics source for M(atrix) theory in 9+1 dimensions. In this section we
want to make this connection to M(atrix) theory directly from SYM$_{10+2}^{1}$
without having to first go through the shadow SYM$_{9+1}^{1}.$ Since this is
the first direct link between 2T-physics and M-theory we want to make the
connection as clear as possible.

The starting point is the action $S_{SYM}$ in Eq.(\ref{action}). Consider at
first the Yang-Mills part\footnote{We are now writing the Yang-Mills group in
matrix version instead of using the adjoint index $a$. The relation between
the two is $A_{M}=A_{M}^{a}t_{a}$ where $t_{a}$ is a hermitian matrix
representation in the fundamental representation of the group $G.$ Then
$\left[  A_{M},A_{N}\right]  =it^{a}\left(  f_{abc}A_{M}^{b}A_{N}^{c}\right)
,$ and $t_{a}$ is normalized as $Tr\left(  t_{a}t_{b}\right)  =2\delta_{ab}.$}
for the case $d+2=12$
\begin{equation}
L_{YM}=-\frac{1}{4g_{YM}^{2}}\sqrt{G}\delta\left(  W\right)  \Omega^{\frac
{3}{2}}\frac{1}{2}Tr\left(  F_{MN}F_{NQ}\right)  G^{MP}G^{NQ}\label{YM2}%
\end{equation}
We split the $10+2$ coordinates $X^{M}$ into two parts $x^{\mu}\sim\left(
9+1\right)  $ and $\sigma^{m}\sim\left(  1+1\right)  $%

\begin{equation}
X^{M}=\left(  \sigma^{m},x^{\mu}\right)  .
\end{equation}
We take all the fields to be independent of $x^{\mu},$ so that they depend
only on $\sigma^{m}.$ Then the Yang-Mills field strength $F_{MN}$ splits into
three parts, $F_{MN}=\left(  F_{mn},F_{m\nu},F_{\mu\nu}\right)  .$ Since all
derivatives with respect to $x^{\mu}$ are dropped, we have (where
$\partial_{m}\equiv\partial/\partial\sigma^{m}$)
\begin{align}
F_{mn}  & =\partial_{m}A_{n}-\partial_{n}A_{m}-i\left[  A_{m},A_{n}\right]
,\\
F_{m\mu}  & =D_{m}A_{\mu}=\partial_{m}A_{\mu}-i\left[  A_{m},A_{\mu}\right]
,\;\;F_{\mu\nu}=-i\left[  A_{\mu},A_{\nu}\right]  ,
\end{align}
We also take $W\left(  \sigma\right)  ,\Omega\left(  \sigma\right)  $ as well
as the metric $G_{MN}\left(  \sigma\right)  $ to be only a function of
$\sigma^{m}$ and of the form
\begin{equation}
G_{MN}=\left(
\begin{array}
[c]{cc}%
g_{mn}\left(  \sigma\right)  & 0\\
0 & a^{2}\left(  \sigma\right)  \eta_{\mu\nu}%
\end{array}
\right)  ,\text{ }\sqrt{G}=\sqrt{g}a^{10}.
\end{equation}
Then, the homothety conditions on the geometry Eqs.(\ref{geom}) are satisfied
with the following forms
\begin{align}
V_{M} &  =\left(  v_{m}\left(  \sigma\right)  ,0\right)  _{M},\;V^{M}%
=(v^{m},0)^{M},\;v_{m}=\frac{1}{2}\partial_{m}W,\;v^{m}=g^{mn}v_{n}.\\
0 &  =\left(  v\cdot\partial-1\right)  a\left(  \sigma\right)  ,\;\left(
v\cdot\partial+4\right)  \Omega\left(  \sigma\right)  ,\;\pounds _{v}%
g_{mn}=2g_{mn},
\end{align}
where $\pounds _{v}$ is the Lie derivative with respect to the two dimensional
vector $v^{m}\left(  \sigma\right)  $ defined above. We see from the last line
that it is consistent to take the warp factor $a\left(  \sigma\right)  $ as a
function of $\Omega\left(  \sigma\right)  $ just as in the previous section
Eq(\ref{aOmeg}), but now as a general function of $\sigma$,%
\begin{equation}
a\left(  \sigma\right)  =\Omega^{-1/4}\left(  \sigma\right)  .\label{aOmeg2}%
\end{equation}
Inserting these forms in the Yang-Mills action, and using $\sqrt{G}%
\Omega^{3/2}=\sqrt{g}\Omega^{3/2}a^{10}=\sqrt{g}a^{4},$ we obtain the
following reduced form%
\begin{align}
L_{YM}  & =-\frac{1}{4g_{YM}^{2}}\delta\left(  W\right)  \sqrt{G}\Omega
^{\frac{3}{2}}\frac{1}{2}Tr\left(  F_{MN}F_{PQ}\right)  G^{MP}G^{NQ}\\
& =-\frac{1}{4g_{YM}^{2}}\delta\left(  W\right)  \sqrt{g}\frac{1}{2}Tr\left\{
\begin{array}
[c]{c}%
a^{4}F_{mn}F^{mn}+2a^{2}\left(  D_{m}A_{\mu}\right)  \left(  D^{m}A^{\mu
}\right) \\
-\left[  A_{\mu},A_{\nu}\right]  \left[  A^{\mu},A^{\nu}\right]
\end{array}
\right\}
\end{align}
where all contractions in $m$ labels are done by using the metric
$g_{mn}\left(  \sigma\right)  $ and in $\mu$ labels by using the Minkowski
metric $\eta_{\mu\nu}.$

Note that the ten fields $A_{\mu}\left(  \sigma\right)  $ behave like scalar
fields as functions of the two dimensional manifold $\sigma^{m}.$ These are
the 10 matrices of M(atrix) theory that are covariant SO$\left(  9,1\right)  $
vectors. Upon variation of the action with respect to the fields, $W\left(
\sigma\right)  ,a\left(  \sigma\right)  $, $g_{mn}\left(  \sigma\right)  $ as
well $A_{\mu}\left(  \sigma\right)  ,$ we derive kinematic and dynamical
equations for each field, as described earlier before Eq.(\ref{geom}). These
can be solved easily, in particular by choosing the two dimensional basis
labelled by $\left(  w,u\right)  $ as it appears as a sub-basis in section
(\ref{technical}). The result of solving the kinematic equations is to produce
the conformal shadow in which the fields shadow in which the fully reduced
M(atrix) theory fields $A_{\mu}$ are now constants independent of $\sigma
^{m},$ whose \textquotedblleft dynamical equations are reproduced by the
shadow action%
\begin{equation}
L_{YM}^{shadow}=\frac{1}{4g_{YM}^{2}}\frac{1}{2}Tr\left(  \left[  A_{\mu
},A_{\nu}\right]  \left[  A^{\mu},A^{\nu}\right]  \right) \label{reduced}%
\end{equation}
This is the bosonic part of the supersymmetric M(atrix) Theory action in the
(-1)-brane version \cite{matrixThJ1}-\cite{matrixThJ4}.

The 0-brane version of \cite{matrixTh1}-\cite{matrixTh2} is derived similarly,
by splitting the coordinates $\left(  10+2\right)  \rightarrow\left(
1+2\right)  \oplus\left(  9+0\right)  \sim\sigma^{m}\oplus x^{i},$ taking all
the fields independent of $x^{i},$ and then following the same procedure as above.

Well before the action in Eq.(\ref{reduced}) was interpreted as M(atrix)
theory during 1996-99, this same action was proposed in 1990 as a bridge
between string theory and large $N$ gauge theory \cite{AreaDiff}. This was
based on the observation that for $N\rightarrow\infty$ one can substitute area
preserving diffeormorphisms for SU$\left(  \infty\right)  .$ In that case the
infinite matrices $\left(  A_{\mu}\right)  _{i}^{j}$ can be expressed in terms
of string coordinates $X_{\mu}\left(  \xi^{\alpha}\right)  $ on the worldsheet
$\xi^{\alpha}\equiv\left(  \tau,\sigma\right)  $, matrix commutators are
reproduced by Poisson brackets $\left[  A_{\mu},A_{\nu}\right]  _{i}%
^{j}\leftrightarrow\left\{  X_{\mu},X_{\nu}\right\}  \left(  \xi^{\alpha
}\right)  =\frac{\partial X_{\mu}}{\partial\tau}\frac{\partial X_{\nu}%
}{\partial s}-\frac{\partial X_{\mu}}{\partial\sigma}\frac{\partial X_{\nu}%
}{\partial\tau},$ and the trace of the infinite matrices is recovered by
integration over the worldsheet $Tr\leftrightarrow\int d^{2}\xi.$ Then the
action in Eq.(\ref{reduced}) is just proportional to $\int d^{2}\xi\det\left(
g\right)  $ where the induced worldsheet metric is
\begin{equation}
g_{\alpha\beta}=\frac{\partial X^{\mu}}{\partial\xi^{\alpha}}\frac{\partial
X^{\nu}}{\partial\xi^{\beta}}\eta_{\mu\nu},\;\det\left(  g\right)  =\left\{
X_{\mu},X_{\nu}\right\}  \left\{  X^{\mu},X^{\nu}\right\}  .\label{poisson}%
\end{equation}
This is a gauge fixed version of the Nambu action
\begin{equation}
\int d^{2}\xi\det\left(  -g\right)  \leftrightarrow\int d^{2}\xi\det\sqrt{-g}%
\end{equation}
where the full diffeomorphism symmetry of the Nambu action has been gauge
fixed to the subgroup of area preserving diffeomorphisms. The results of the
present paper now show that all of this is recovered from the dimensional
reduction of SYM$_{10+2}^{1}$.

We now turn to the fermionic terms. Starting from SYM$_{10+2}^{1}$ in
Eq.(\ref{action}), after replacing the adjoint index $a$ by matrices as
above,
\begin{equation}
L_{SYM}^{fermi}=\frac{i}{2}\sqrt{-G}\delta\left(  W\left(  X\right)  \right)
\frac{1}{2}Tr\left[  \overline{\lambda}V\bar{D}\lambda+\overline{\lambda
}\overleftarrow{D}\overline{V}\lambda\right]  ,
\end{equation}
we follow the same procedure of dimensional reduction. We again have,
$\sqrt{-G}\delta\left(  W\left(  X\right)  \right)  =\sqrt{-g}a^{10}%
\delta\left(  W\left(  \sigma\right)  \right)  .$ Also, because $V^{M}$ and
$\partial_{M}$ are vanishing when $M=\mu$, we get
\begin{align}
& \delta\left(  W\left(  X\right)  \right)  \sqrt{-G}\overline{\lambda}%
V\bar{D}\lambda\nonumber\\
& =\delta\left(  W\left(  \sigma\right)  \right)  \sqrt{-g}a^{10}%
\overline{\lambda}\left(  v_{n}\Gamma^{n}\right)  \left\{  \bar{\Gamma}%
^{m}D_{m}\lambda+\bar{\Gamma}^{\mu}\left(  \frac{1}{4}\omega_{\mu}^{ij}%
\Gamma_{ij}\lambda-i\left[  A_{\mu},\lambda\right]  \right)  \right\}
.\label{spincontrib}%
\end{align}
The spin connection $\omega_{M}^{ij}$ that is compatible with the Sp$\left(
2,R\right)  $ conditions in (\ref{geom}), namely $E_{M}^{i}=D_{M}V^{i}$ has
only the following non-zero components (with $i=\hat{m}\oplus\hat{\mu}$ the
tangent indices)
\begin{equation}
\omega_{M}^{ij}\left(  x,y\right)  =%
\begin{tabular}
[c]{|c|c|c|c|}\hline
$_{M}$
%TCIMACRO{\TEXTsymbol{\backslash}}%
%BeginExpansion
$\backslash$%
%EndExpansion
$~~ij$ & $^{\hat{m}\hat{n}}$ & $^{\hat{m}\hat{\mu}}$ & $^{\hat{\mu}\hat{\nu}}%
$\\\hline
$\omega_{m}^{ij}=$ & $\;\;\;\;\;\omega_{m}^{\hat{m}\hat{n}}\;\;\;\;\;$ & $0 $
& $~~~0~~$\\\hline
$\omega_{\mu}^{ij}=$ & $0$ & $\omega_{\mu}^{\hat{m}\hat{\mu}}=\delta_{\mu
}^{\hat{\mu}}\delta^{\hat{m}n}\partial_{n}\ln a$ & $~~~0~~$\\\hline
\end{tabular}
\end{equation}
The covariant derivative $D_{m}$ includes $\omega_{m}^{\hat{m}\hat{n}},$ which
is the standard spin connection constructed from a vielbein. The contribution
from $\omega_{\mu}^{\hat{m}\hat{\mu}}$ in Eq.(\ref{spincontrib}) comes in the
form
\begin{equation}
\frac{2}{4}\omega_{\mu}^{\hat{m}\hat{\mu}}\bar{\Gamma}^{\mu}\Gamma_{\hat{m}%
}\bar{\Gamma}_{\hat{\mu}}=\frac{1}{2}\delta_{\mu}^{\hat{\mu}}\delta^{\hat{m}%
n}\partial_{n}\ln a\left(  -\Gamma_{\hat{m}}\Gamma^{\mu}\bar{\Gamma}_{\hat
{\mu}}\right)  =-\frac{10}{2}\bar{\Gamma}^{n}\partial_{n}\ln a
\end{equation}
This is just right to absorb all dependence on the warp factor $a\left(
\sigma\right)  $ into a rescaling of the fermion, as follows%
\begin{equation}
\left(  a^{5}\overline{\lambda}\right)  \left(  v_{n}\Gamma^{n}\right)
\left\{  \bar{\Gamma}^{m}D_{m}\left(  \lambda a^{5}\right)  +\bar{\Gamma}%
^{\mu}\left(  -i\left[  A_{\mu},\left(  \lambda a^{5}\right)  \right]
\right)  \right\}
\end{equation}
Thus, by defining a renormalized 32-spinor given by%
\begin{equation}
\psi\equiv\left(  \lambda a^{5}\right)  \sqrt{4g_{YM}}%
\end{equation}
we manage to write the fermion action in the form%
\begin{equation}
L_{SYM}^{fermi}=\frac{1}{4g_{YM}^{2}}\delta\left(  W\left(  \sigma\right)
\right)  \sqrt{-g}\frac{1}{2}Tr\left\{  \frac{1}{2}\left(  i\bar{\psi}%
v\bar{\Gamma}^{m}D_{m}\psi+h.c.\right)  +\bar{\psi}v\bar{\Gamma}^{\mu}\left[
A_{\mu},\psi\right]  \right\}
\end{equation}
Hence, the total reduced action for SYM$_{10+2}^{1}$ is%
\begin{equation}
S_{SYM}^{reduced}=\frac{1}{8g_{YM}^{2}}\int d^{2}\sigma\delta\left(  W\left(
\sigma\right)  \right)  \sqrt{-g}Tr\left\{
\begin{array}
[c]{c}%
-a^{4}F_{mn}F^{mn}-2a^{2}\left(  D_{m}A_{\mu}\right)  \left(  D^{m}A^{\mu
}\right) \\
+\frac{1}{2}\left(  i\bar{\psi}v\bar{\Gamma}^{m}D_{m}\psi+h.c.\right) \\
+\left[  A_{\mu},A_{\nu}\right]  \left[  A^{\mu},A^{\nu}\right]  +\bar{\psi
}v\bar{\Gamma}^{\mu}\left[  A_{\mu},\psi\right]
\end{array}
\right\} \label{reducedTot}%
\end{equation}
Note that here the field $\lambda\left(  \sigma\right)  $ has 32 real
components, but there is a kappa-type local symmetry, as in all 2T-physics
actions that involve fermions \cite{2tstandardM}, that eliminates half of the
fermions by a gauge choice, thus really only 16 real fermion degrees of
freedom are present. This is just the right content in M(atrix) theory.

As outlined just before Eq.(\ref{reduced}), solving the kinematic equations
derived from this action produces the shadow which is recognized as the
supersymmetrized M(atrix) theory that generalizes Eq.(\ref{reduced}), with
matrices $\left(  A_{\mu}\right)  _{i}^{j}$ and $\left(  \psi_{+}\right)
_{i}^{j}$ that are independent of the two coordinates $\sigma^{m}$
\begin{equation}
L_{SYM}^{shadow}=\frac{1}{8g_{YM}^{2}}Tr\left\{  \left[  A_{\mu},A_{\nu
}\right]  \left[  A^{\mu},A^{\nu}\right]  +\bar{\psi}_{+}\bar{\gamma}^{\mu
}\left[  A_{\mu},\psi_{+}\right]  \right\}  .
\end{equation}
Here $\psi_{+}$ is the 16-component spinor of SO$\left(  9,1\right)  $ that
corresponds to half of the 32-component $\psi.$ Before choosing gauges or
solving the kinematic equations, the 32 components $\psi$ is a reminder and a
link to 10+2 dimensions.

For large $N$ this action (as well as its parent in Eq.(\ref{reducedTot})) may
be rewritten in terms of Poisson brackets on a worldsheet \cite{AreaDiff} as
in Eq.(\ref{poisson}).

By using similar methods, other versions of M(atrix) theory that relate to
0-branes, 1-branes, and more generally p-branes \cite{matrixTh1}%
-\cite{matrixTh2} can be derived directly from the action of SYM$_{10+2}^{1}$
in Eq.(\ref{action}) by various dimensional reductions or compactifications
that parallel those in \cite{matrixTh1}-\cite{matrixTh2}.

\section{Closing comments}

Having established that the conventional 1T-physics methods miss
systematically a vast amount of information even in simple classical or
quantum mechanics (see recent summary \cite{phaseSpace} and the introduction
in \cite{2tScalars}) it is reasonable to expect that progress in fundamental
physics, in particular the quest for the fundamental principles, would benefit
from the methods of 2T-physics. It is with this in mind that we have embarked
on constructing the higher dimensional 2T theories that connect to well known
and cherished theories in 1T-physics. In this paper we have discussed the
first such theory in 10+2 dimensions, a number of dimensions that was not
reached before, and have shown that it is the source, and unifying factor, of
well known lower dimensional theories.

The process of derivation is a combination of dimensional reduction and
extracting a shadow of 2T-physics by solving the kinematic equations that
follow from the 2T action. The kinematic equations amount to imposing the
gauge symmetry requirements of Sp$\left(  2,R\right)  $ in phase space, as
summarized recently in \cite{phaseSpace}\cite{2tScalars}. In principle there
are many other types of shadows and compactifications derivable from 2T-field
theory that can lead to other dual versions of each of the theories discussed
here, as sketched in Fig.1. The additional shadows produced by 2T field theory
have so far been little explored in the context of field theory
\cite{emergentfieldth1}\cite{emergentfieldth2} although they are much better
developed in the context of classical or quantum particle mechanics
\cite{phaseSpace}.

By using the web of connections that we discussed here, and those that can be
further derived, one can in principle establish a web of dualities or
connections among various 1T-theories that were not suspected before. This
additional predicted information, which can be verified in 1T-physics, is
related to the extra dimensions as is already captured by the same unifying
theory in 10+2 dimensions. Hence, studying directly the theory in 10+2
dimensions (for example as in \cite{emergentfieldth1}\cite{emergentfieldth2}%
\cite{rivelles}\cite{weinberg}) can yield many benefits and predictions for
the lower dimensional theories. In addition to the deeper implications that
our program has about the meaning of space-time, exploring the hidden
symmetries and dualities related to the 10+2 dimensional parent SYM$_{10+2}%
^{1}$ theory is expected to yield many practical side benefits, including new
computational techniques that could clarify or supplement those already used
in SYM$_{3+1}^{4}$ and in M(atrix) theory.

The path of research pursued in the current paper is expected to lead to
supergravity in 10+2 and 11+2 dimensions and eventually to a 2T approach to
M-theory and its dualities. This should provide a dynamical and gauge symmetry
basis for F-theory \cite{ftheory} and S-theory \cite{Stheory} from deeper
phase space gauge symmetry principles \cite{phaseSpace}\cite{noncommutative}
which require higher spacetime with two times.

\newpage

\appendix{}

\section{Conformally flat shadow spacetimes in $d$ dimensions from flat $d+2$
spacetime \label{flat}}

The topic of this appendix was part of the discussion in
\cite{emergentfieldth1}\cite{emergentfieldth2} on the shadows of 2T field
theory in flat spacetime. But in this appendix we present a more systematic
approach for the conformal shadow, including the expansion in powers of $w$
that was not covered in \cite{emergentfieldth1}\cite{emergentfieldth2}.

Consider the line element in flat spacetime in $d+2$ dimensions parametrized
as
\begin{align}
ds_{d+2}^{2} &  =dX^{i}dX^{j}\eta_{ij}=-\left(  dX^{0^{\prime}}\right)
^{2}+\left(  dX^{1^{\prime}}\right)  ^{2}+dX^{\alpha}dX^{a}\eta_{ab}\\
&  =-2dX^{+^{\prime}}dX^{-^{\prime}}+dX^{\alpha}dX^{a}\eta_{ab}~,
\end{align}
where $\eta_{ij}$ is the flat metric with SO$\left(  d,2\right)  $ symmetry
and $\eta_{ab}$ is the Minkowski metric with SO$\left(  d-1,1\right)  $
symmetry. We parametrize these flat Cartesian coordinates $X^{i}$ , with
$i=\left(  \pm^{\prime},a\right)  $ labeling the flat basis, in terms of
curvilinear coordinates $X^{M}=\left(  w,u,x^{\mu}\right)  ^{M},$ where $M$
labels the curvilinear basis (hence compared to the curved basis in the text),
as follows%

\begin{align}
X^{+^{\prime}} &  =\frac{X^{0^{\prime}}+X^{1^{\prime}}}{\sqrt{2}}=\pm
e^{-2\Sigma},\ \ \;X^{a}=e^{-2\Sigma}q^{a},\\
X^{-^{\prime}} &  =\frac{X^{0^{\prime}}-X^{1^{\prime}}}{\sqrt{2}}=\pm
e^{-2\Sigma}\frac{q^{2}}{2}\mp e^{2\Sigma}\frac{w}{2},
\end{align}
where $\Sigma$ and $q^{a}$ are arbitrary functions of the curvilinear
coordinates $\left(  w,u,x\right)  .$ The point of this parametrization is
that computing $X^{2}=X^{i}X^{j}\eta_{ij}$ we find $X^{2}=w.$ That is, $X^{2}$
as computed with the flat Cartesian coordinates $X^{i}$ coincides with the
curvilinear coordinate $w.$ After computing $dX^{i}$ and inserting in
$ds_{d+2}^{2}=dX^{i}dX^{j}\eta_{ij}$ the flat metric above takes the form%
\begin{equation}
ds_{d+2}^{2}=-2dw\left(  d\Sigma\right)  -4w\left(  d\Sigma\right)
^{2}+e^{-4\Sigma}\left(  dq\right)  ^{2}.
\end{equation}
We will take the following specialized form for $\Sigma\left(  w,u,x\right)  $
and $q^{a}\left(  w,u,x\right)  $
\begin{equation}
\Sigma\left(  w,u,x\right)  =u+\frac{1}{2}\sigma\left(  x,we^{4u}\right)
,\;\;q^{a}\left(  w,u,x\right)  =q^{a}\left(  x,we^{4u}\right)  .
\end{equation}
This form is motivated by previous work \cite{emergentfieldth1}%
\cite{emergentfieldth2}\cite{2tGravity}\cite{2tGravDetails} which shows the
relevance of the combination of coordinates $we^{4u}\equiv z$. This gives%
\[
ds_{d+2}^{2}=\left\{
\begin{array}
[c]{l}%
-\left(  dw\right)  ^{2}\left[  \sigma^{\prime}\left(  1+we^{4u}\sigma
^{\prime}\right)  -e^{-2\sigma}\left(  q^{\prime}\right)  ^{2}\right]
e^{4u}\\
+\left(  du\right)  ^{2}\left[  -4w+16\left(  we^{4u}\right)  ^{2}\left[
\sigma^{\prime}\left(  1+we^{4u}\sigma^{\prime}\right)  -e^{-2\sigma}\left(
q^{\prime}\right)  ^{2}\right]  \right] \\
+dx^{\mu}dx^{\nu}\left[  -\left(  we^{4u}\right)  \partial_{\mu}\sigma
\partial_{\nu}\sigma+e^{-2\sigma}\partial_{\mu}q\cdot\partial_{\nu}q\right]
e^{-4u}\\
+2dwdu\left[  -1-4z\left(  \sigma^{\prime}\left(  1+we^{4u}\sigma^{\prime
}\right)  -e^{-2\sigma}\left(  q^{\prime}\right)  ^{2}\right)  \right] \\
+2dwdx^{\mu}\left[  -\left(  \frac{1}{2}+we^{4u}\sigma^{\prime}\right)
\partial_{\mu}\sigma+e^{-2\sigma}\partial_{\mu}q\cdot q^{\prime}\right] \\
+2dudx^{\mu}\left[  -\left(  \frac{1}{2}+we^{4u}\sigma^{\prime}\right)
\partial_{\mu}\sigma+e^{-2\sigma}\partial_{\mu}q\cdot q^{\prime}\right]  4w
\end{array}
\right\}  .
\]
where $\sigma^{\prime}$ and $q_{a}^{\prime}$ are defined as the \textit{total}
derivatives with respect to the variable $z\equiv we^{4u}$
\begin{equation}
q_{a}^{\prime}\equiv\frac{dq^{a}\left(  x,z\right)  }{dz},\;\;\sigma^{\prime
}\equiv\frac{d\sigma\left(  q^{a}\left(  x,z\right)  ,z\right)  }{dz}%
=\frac{\partial\sigma}{\partial z}+\frac{\partial\sigma}{\partial q^{a}}%
q_{a}^{\prime}.
\end{equation}
It was argued in \cite{2tGravity}\cite{2tGravDetails} that a general metric in
2T-gravity (therefore in particular the flat case in this Appendix) can be
brought to the following standard gauge fixed form which is appropriate for
the conformal shadow and its prolongations \cite{2tGravDetails} at any $w $%
\begin{equation}
ds^{2}=-2dwdu-4w\left(  du\right)  ^{2}+e^{-4u}g_{\mu\nu}\left(
x,we^{4u}\right)  dx^{\mu}dx^{\nu}.
\end{equation}
If the above expression for $ds_{d+2}^{2}$ is to agree with this gauge fixed
form we must put further constraints on $\sigma,q^{a}$ as follows%
\begin{align}
e^{-2\sigma}\left(  q^{\prime}\right)  ^{2} &  =\sigma^{\prime}\left(
1+we^{4u}\sigma^{\prime}\right)  ,\\
2\partial_{\mu}q\cdot q^{\prime} &  =\left(  1+2we^{4u}\sigma^{\prime}\right)
e^{2\sigma}\partial_{\mu}\sigma.
\end{align}
These equations are solved uniquely by the following expressions for
$\sigma^{\prime},q_{a}^{\prime}$ (using the chain rule $\partial_{\mu}%
\sigma=\frac{\partial q^{a}}{\partial x^{\mu}}\frac{\partial\sigma}{\partial
q^{a}}$)
\begin{align}
q_{a}^{\prime} &  =\frac{e^{2\sigma}\partial_{a}\sigma}{2\sqrt{1-ze^{2\sigma
}\left(  \partial_{a}\sigma\right)  ^{2}}},\\
\sigma^{\prime} &  =\frac{e^{2\sigma}\left(  \partial_{a}\sigma\right)  ^{2}%
}{2\sqrt{1-ze^{2\sigma}\left(  \partial_{a}\sigma\right)  ^{2}}\left(
1+\sqrt{1-ze^{2\sigma}\left(  \partial_{a}\sigma\right)  ^{2}}\right)
}.\label{sigmaEq}%
\end{align}
where $\partial_{a}\sigma\equiv\frac{\partial\sigma}{\partial q^{a}},$ and
$\left(  \partial_{a}\sigma\right)  ^{2}\equiv\eta^{ab}\partial_{a}%
\sigma\partial_{b}\sigma.$ We see that (\ref{sigmaEq}) is a partial
differential equation for $\sigma\left(  q^{a},z\right)  $ as a function of
$d+1$ coordinates $q^{a},z$. Once $\sigma\left(  q^{a},z\right)  $ is
determined by solving this equation, we can find $q_{a}\left(  x,z\right)  $
by integrating the first equation with respect to $z$
\begin{equation}
q^{a}\left(  x,z\right)  =q_{0}^{a}\left(  x\right)  +\int_{0}^{z}dz^{\prime
}\frac{e^{2\sigma}\partial_{b}\sigma\eta^{ab}}{2\sqrt{1-z^{\prime}e^{2\sigma
}\left(  \partial_{a}\sigma\right)  ^{2}}}.
\end{equation}
where $q_{0}^{a}\left(  x\right)  $ is completely arbitrary.

There remains solving the $\sigma$-equation (\ref{sigmaEq}). It is useful to
do this by expanding both $q^{a}\left(  x,z\right)  $ and $\sigma\left(
q^{a}\left(  x,z\right)  ,z\right)  $ in powers of $z(=we^{4u})$ since after
all we are only interested in the first few powers in $z$ on account of the
delta function $\delta\left(  w\right)  $ in the action (\ref{action1}). So,
we define the expansion%
\begin{align}
q^{a}\left(  x,z\right)   &  =q_{0}^{a}\left(  x\right)  +zq_{1}^{a}\left(
x\right)  +\frac{z^{2}}{2}q_{2}^{a}\left(  x\right)  +\cdots\\
\sigma\left(  q^{a}\left(  x,z\right)  ,z\right)   &  =\sigma_{0}\left(
q_{0}\left(  x\right)  \right)  +z\sigma_{1}\left(  q_{0}\left(  x\right)
\right)  +\frac{z^{2}}{2}\sigma_{2}\left(  q_{0}\left(  x\right)  \right)
+\cdots
\end{align}
By inserting these back into the $\sigma$-equation (\ref{sigmaEq}) and noting
that we can use $\partial_{a}\sigma=\frac{\partial\sigma}{\partial q^{a}%
}=\frac{\partial\sigma}{\partial q_{0}^{a}},$ we easily obtain an explicit
solution for $\sigma_{1},\sigma_{2},\cdots,$ and $q_{1}^{a},q_{2}^{a},\cdots$
in terms of the arbitrary $d+1$ functions of spacetime $q_{0}^{a}\left(
x\right)  ,\sigma_{0}\left(  q_{0}\left(  x\right)  \right)  .$ The result
looks as follows up to $O\left(  z^{2}\right)  $
\begin{equation}%
\begin{array}
[c]{c}%
q^{a}\left(  x,z\right)  =q_{0}^{a}\left(  x\right)  +ze^{2\sigma_{0}}%
\frac{\partial\sigma_{0}\left(  q_{0}\right)  }{2\partial q_{0}^{a}}+\cdots\\
\sigma\left(  x,z\right)  =\sigma_{0}\left(  q_{0}\right)  +ze^{2\sigma_{0}%
}\left(  \frac{\partial\sigma_{0}\left(  q_{0}\right)  }{2\partial q_{0}^{a}%
}\right)  ^{2}+\cdots\\
q_{0}^{a}\left(  x\right)  ,\sigma_{0}\left(  q_{0}\left(  x\right)  \right)
\text{ are arbitrary.}%
\end{array}
\label{solutions}%
\end{equation}
It is evident that the coefficients of all higher powers in $z$ in both
$\sigma\left(  x,z\right)  $ and $q^{a}\left(  x,z\right)  $ are completely
fixed by the $d+1$ arbitrary functions $q_{0}^{a}\left(  x\right)  ,\sigma
_{0}\left(  q_{0}\left(  x\right)  \right)  =\sigma_{0}\left(  x\right)  .$

With this result we now analyze again the line element which now has the
standard gauge fixed form (for the conformal shadow)
\begin{equation}
ds^{2}=-2dwdu-4w\left(  du\right)  ^{2}+e^{-4u}g_{\mu\nu}\left(
x,we^{4u}\right)  dx^{\mu}dx^{\nu},
\end{equation}
and find that $g_{\mu\nu}\left(  x,z\right)  $ is given by
\begin{align}
g_{\mu\nu}\left(  x,z\right)   &  =e^{-2\sigma}\partial_{\mu}q^{a}%
\partial_{\nu}q^{b}\eta_{ab}-z\partial_{\mu}\sigma\partial_{\nu}\sigma\\
&  =g_{\mu\nu}^{\left(  0\right)  }\left(  x\right)  +zg_{\mu\nu}^{\left(
1\right)  }\left(  x\right)  +\frac{z^{2}}{2}g_{\mu\nu}^{\left(  2\right)
}\left(  x\right)  +\cdots
\end{align}
By inserting our solutions for $q^{a}\left(  x,z\right)  $ and $\sigma\left(
x,z\right)  $ we compute $g_{\mu\nu}^{\left(  0\right)  },g_{\mu\nu}^{\left(
1\right)  },g_{\mu\nu}^{\left(  2\right)  },\cdots$ as follows
\begin{align}
g_{\mu\nu}^{\left(  0\right)  }\left(  x\right)   &  =e_{\mu}^{a}\left(
x\right)  e_{\nu}^{b}\left(  x\right)  \eta_{ab},\;\;\text{where }e_{\mu}%
^{a}\left(  x\right)  =e^{-\sigma_{0}\left(  x\right)  }\frac{\partial
q_{0}^{a}\left(  x\right)  }{\partial x^{\mu}},\label{vielbein0}\\
g_{\mu\nu}^{\left(  1\right)  }\left(  x\right)   &  =-\frac{1}{2}\left(
\partial_{a}\sigma_{0}\right)  ^{2}\partial_{\mu}q_{0}\cdot\partial_{\nu}%
q_{0}+\partial_{\mu}\sigma_{0}\partial_{\nu}\sigma_{0}+\left(  \partial_{\mu
}q_{0}^{b}\partial_{\nu}q_{0}^{a}\right)  \partial_{b}\partial_{a}\sigma
_{0},\\
g_{\mu\nu}^{\left(  2\right)  }\left(  x\right)   &  =\cdots
\end{align}
In the expression for $g_{\mu\nu}^{\left(  1\right)  }\left(  x\right)  ,$
assuming that $\sigma_{0}\left(  q_{0}\left(  x\right)  \right)  =\sigma
_{0}\left(  x\right)  $ is chosen as a function of $x^{\mu},$ we can evaluate
the derivatives $\partial_{a}\sigma_{0}$ by using the chain rule
\begin{equation}
\partial_{a}\sigma_{0}\left(  x\left(  q_{0}\right)  \right)  =\frac{\partial
x^{\mu}}{\partial q_{0}^{a}}\partial_{\mu}\sigma_{0}\left(  x\right)
=e^{-\sigma_{0}\left(  x\right)  }e_{a}^{\mu}\left(  x\right)  \partial_{\mu
}\sigma_{0}\left(  x\right)  ,
\end{equation}
where $e_{a}^{\mu}\left(  x\right)  $ is the inverse of the vielbein defined
in (\ref{vielbein0}). So, the simple rule is $\partial_{a}\left(
e^{\sigma_{0}}\right)  =e_{a}^{\mu}\left(  x\right)  \partial_{\mu}\sigma
_{0}\left(  x\right)  .$

The lowest component $g_{\mu\nu}^{\left(  0\right)  }\left(  x\right)
=e_{\mu}^{a}\left(  x\right)  e_{\nu}^{b}\left(  x\right)  \eta_{ab}$ alone
determines the geometric properties of the shadow in $d$ dimensions
\cite{2tGravDetails}, and from the form of the vielbein $e_{\mu}^{a}\left(
x\right)  =e^{-\sigma_{0}\left(  x\right)  }\frac{\partial q_{0}^{a}\left(
x\right)  }{\partial x^{\mu}}$ we see that the spacetime of the shadow is a
conformally flat spacetime.

The higher components of the metric $g_{\mu\nu}^{\left(  1\right)  }\left(
x\right)  ,g_{\mu\nu}^{\left(  2\right)  }\left(  x\right)  ,\cdots$ determine
the geometric properties of the \textit{prolongations} of the shadow as
discussed in \cite{2tGravDetails}, but these do not interfere with the self
consistent 1-time physics of the shadow in the spacetime given by $g_{\mu\nu
}^{\left(  0\right)  }\left(  x\right)  $ \cite{2tGravDetails}.

\section{More general solution of the SUSY condition \label{Scondition}}

In this appendix we find a more general solution of the SUSY condition
(\ref{SUSYcond}) which is regarded as a constraint on the SUSY parameter
$\varepsilon_{A}$
\begin{equation}
\left[  -\frac{d-4}{d-2}\left(  \bar{\Gamma}^{PQN}\Gamma^{M}\varepsilon
\right)  _{A}V_{N}\partial_{M}\ln\Omega+\left(  \bar{\Gamma}^{M}\Gamma
^{PQN}D_{M}\varepsilon\right)  _{A}V_{N}\right]  _{W=0}=\left[  V^{P}U_{A}%
^{Q}-V^{Q}U_{A}^{P}\right]  _{W=0}.
\end{equation}
Recall that the $U_{A}^{P}$ are arbitrary. In the $X^{M}=\left(  w,u,x^{\mu
}\right)  ^{M}$ basis, appropriate for the conformal shadow, we use the
results in Eqs.(\ref{Vs}-\ref{omeps}), in particular $V_{N}=\left(  \frac
{1}{2},0,0\right)  _{N}$ and $V^{N}=\left(  2w,-\frac{1}{2},0\right)  ^{N}$
and then set $W\left(  X\right)  =w=0$ to simplify this expression
\begin{equation}
\left[  -\frac{d-4}{d-2}\left(  \bar{\Gamma}^{PQw}\Gamma^{M}\varepsilon
\right)  _{A}\partial_{M}\ln\Omega+\left(  \bar{\Gamma}^{M}\Gamma^{PQw}%
D_{M}\varepsilon\right)  _{A}=-\left(  \delta_{u}^{P}U_{A}^{Q}-\delta_{u}%
^{Q}U_{A}^{P}\right)  \right]  _{w=0}.
\end{equation}
We will also use the expressions for the spin connection $\omega_{M}^{ij}$
given in (\ref{spinCo},\ref{spinComu}) at $w=0$ (after derivatives
$\partial_{w}$ are taken). Next we specialize the antisymmetric indices
$\left[  PQ\right]  $ to examine systematically the various tensor components
of the SUSY condition as follows. For $PQ=wu$ we have
\begin{equation}
\left[  -\frac{d-4}{d-2}\left(  \bar{\Gamma}^{wuw}\Gamma^{M}\varepsilon
\right)  _{A}\partial_{M}\ln\Omega+\left(  \bar{\Gamma}^{M}\Gamma^{wuw}%
D_{M}\varepsilon\right)  _{A}=-\left(  \delta_{u}^{w}V^{w}U_{A}^{u}-\delta
_{u}^{u}U_{A}^{w}\right)  \right]  _{w=0}.
\end{equation}
Noting that at $w=0,$ we get $\bar{\Gamma}^{wuw}=\Gamma^{wuw}=\delta_{u}%
^{w}=0$ and $\delta_{u}^{u}=1,$ this equation determines $U_{A}^{w}$
\begin{equation}
U_{A}^{w}=0.
\end{equation}
Next we take $PQ=w\lambda$
\begin{equation}
\left[  -\frac{d-4}{d-2}\left(  \bar{\Gamma}^{w\lambda w}\Gamma^{M}%
\varepsilon\right)  _{A}\partial_{M}\ln\Omega+\left(  \bar{\Gamma}^{M}%
\Gamma^{w\lambda w}D_{M}\varepsilon\right)  _{A}=-\left(  \delta_{u}^{w}%
U_{A}^{\lambda}-\delta_{u}^{\lambda}U_{A}^{w}\right)  \right]  _{w=0}.
\end{equation}
Every term on both sides of this equation vanishes since $\Gamma^{w\lambda
w}=\delta_{u}^{w}=\delta_{u}^{\lambda}=0$, so this is an identity. Next we
take $PQ=u\lambda$
\[
\left[  -\frac{d-4}{d-2}\left(  \bar{\Gamma}^{u\lambda w}\Gamma^{M}%
\varepsilon\right)  _{A}\partial_{M}\ln\Omega+\left(  \bar{\Gamma}^{M}%
\Gamma^{u\lambda w}D_{M}\varepsilon\right)  _{A}=-\left(  \delta_{u}^{u}%
U_{A}^{\lambda}-\delta_{u}^{\lambda}U_{A}^{u}\right)  \right]  _{w=0}.
\]
We use $\bar{\Gamma}^{u\lambda w}=\bar{\Gamma}^{wu}\bar{\Gamma}^{\lambda}$ and
$\Gamma^{u\lambda w}=\Gamma^{wu}\Gamma^{\lambda}$ to determine $U_{A}%
^{\lambda}$ in the form
\begin{equation}
U^{\lambda}=-\bar{\Gamma}^{wu}\left[  -\frac{d-4}{d-2}\bar{\Gamma}^{\lambda
}\Gamma^{M}\varepsilon\partial_{M}\ln\Omega+\left(  \tilde{\Gamma}^{M}%
\Gamma^{\lambda}D_{M}\varepsilon\right)  \right]  _{w=0}%
\end{equation}
where $\tilde{\Gamma}^{M}=\bar{\Gamma}^{wu}\bar{\Gamma}^{M}\bar{\Gamma}%
^{wu}=\left(  -\Gamma^{w},-\Gamma^{u},\Gamma^{\mu}\right)  ^{M}.$ In this
expression we are supposed to insert $\Omega\left(  X\right)  =e^{\left(
d-2\right)  u}\hat{\Omega}\left(  x,we^{4u}\right)  \;$and$\;\varepsilon
\left(  X\right)  =\exp\left(  u\Gamma^{+^{\prime}-^{\prime}}\right)
\hat{\varepsilon}\left(  x,we^{4u}\right)  $ as determined (\ref{omeps}) and
set $w=0.$

So far there has been no conditions on $\hat{\varepsilon}_{A},$ but the next
case of $PQ=\nu\lambda$ produces conditions on $\hat{\varepsilon}_{A}$ as
follows
\[
\left[  -\frac{d-4}{d-2}\left(  \bar{\Gamma}^{\nu\lambda w}\Gamma
^{M}\varepsilon\right)  \partial_{M}\ln\Omega+\left(  \bar{\Gamma}^{M}%
\Gamma^{\nu\lambda w}D_{M}\varepsilon\right)  =-\left(  \delta_{u}^{\nu}%
U_{A}^{\lambda}-\delta_{u}^{\lambda}U_{A}^{\nu}\right)  \right]  _{w=0}.
\]
After using $\bar{\Gamma}^{\nu\lambda w}=\bar{\Gamma}^{\nu\lambda}\bar{\Gamma
}^{w}$ and noting that the right hand side vanishes on account of $\delta
_{u}^{\nu}=0,$ we get
\begin{equation}
\left[  -\frac{d-4}{d-2}\left(  \bar{\Gamma}^{\nu\lambda}\bar{\Gamma}%
^{w}\Gamma^{M}\varepsilon\right)  \partial_{M}\ln\Omega+\left(  \bar{\Gamma
}^{M}\Gamma^{\nu\lambda}\Gamma^{w}D_{M}\varepsilon\right)  \right]  _{w=0}=0.
\end{equation}
In this expression both $M=w$ terms drop because $\left[  \bar{\Gamma}%
^{w}\Gamma^{w}\right]  _{w=0}=\left[  G^{ww}\right]  _{w=0}=0.$ Furthermore,
the term $D_{u}\varepsilon$ drops because $\left[  D_{u}\varepsilon\right]
_{w=0}=0$ as in (\ref{omeps}), and we can set $\partial_{u}\ln\Omega=d-2$ at
$w=0$ on account of (\ref{omeps}). The result has the form
\begin{equation}
0=\left\{  \bar{\Gamma}^{w}\Gamma^{\nu\lambda}\left[  -\left(  d-4\right)
\Gamma^{u}\varepsilon-\frac{d-4}{d-2}\left(  \partial_{\mu}\ln\Omega\right)
\Gamma^{\mu}\varepsilon\right]  -\bar{\Gamma}^{w}\Gamma^{\mu}\bar{\Gamma}%
^{\nu\lambda}D_{\mu}\varepsilon\right\}  _{w=0}.
\end{equation}

As an example consider $d+2=12$ (for the other cases $d+2=5,6,8$ the
discussion is similar, by changing only the size of the spinor). Since in
12-dimensions $\left[  \bar{\Gamma}^{w}\right]  _{w=0}$ is a 32$\times32$
matrix proportional to $\left(
%TCIMACRO{\QATOP{0}{1}}%
%BeginExpansion
\genfrac{}{}{0pt}{}{0}{1}%
%EndExpansion%
%TCIMACRO{\QATOP{0}{0}}%
%BeginExpansion
\genfrac{}{}{0pt}{}{0}{0}%
%EndExpansion
\right)  $ where each entry is a 16$\times16$ matrix, this equation amounts to
16 equations imposed on the 32 components of $\varepsilon_{A}.$ Taking into
account (\ref{omeps}) we write the 32 component $\varepsilon_{A}\left(
X\right)  $ in terms of two 16-component pieces $\varepsilon_{1}\left(
x\right)  ,\varepsilon_{2}\left(  x\right)  $ at $w=0$
\begin{equation}
\left[  \varepsilon_{A}\left(  X\right)  \right]  _{w=0}=\left(
\begin{array}
[c]{c}%
e^{-u}\varepsilon_{1}\left(  x\right) \\
e^{+u}\varepsilon_{2}\left(  x\right)
\end{array}
\right) \label{e1e2}%
\end{equation}
Then the 16 equations above take the form\footnote{In arriving at this
expression we have used the 2T form of the spin connection in (\ref{spinComu})
to evaluate $\left(  D_{\mu}\varepsilon\right)  _{A}=\left(  \partial_{\mu
}+\frac{1}{4}\omega_{\mu}^{ab}\bar{\Gamma}_{ab}+\frac{1}{2}\omega_{\mu
}^{+^{\prime}b}\bar{\Gamma}_{b}\Gamma^{-^{\prime}}+\frac{1}{2}\omega_{\mu
}^{-^{\prime}b}\bar{\Gamma}_{b}\Gamma^{+^{\prime}}\right)  \varepsilon_{A},$}
\[
0=-\left(  d-4\right)  \left(  -i\sqrt{2}\right)  \bar{\gamma}^{\nu\lambda
}\varepsilon_{2}-\partial_{\mu}\ln\phi^{\frac{d-4}{d-2}}\bar{\gamma}%
^{\nu\lambda}\bar{\gamma}^{\mu}\varepsilon_{1}-\bar{\gamma}^{\mu}\gamma
^{\nu\lambda}\left(  D_{\mu}\varepsilon_{1}+i\sqrt{2}\gamma_{\mu}%
\varepsilon_{2}\right)  ,\;
\]
where in the expression for $D_{\mu}\varepsilon_{1}$ only the usual 1T form of
the spin connection $\omega_{\mu}^{ab}$ appears. Now we use $\bar{\gamma}%
^{\mu}\gamma^{\nu\lambda}\gamma_{\mu}=\left(  d-4\right)  \gamma^{\nu\lambda}$
and notice that $\varepsilon_{2}$ drops out of this equation, so the
constraint on $\varepsilon$ simplifies to a constraint only on $\varepsilon
_{1}\left(  x\right)  $%
\begin{equation}
-\bar{\gamma}^{\nu\lambda}\bar{\gamma}^{\mu}\varepsilon_{1}\left(
\partial_{\mu}\ln\phi^{\frac{d-4}{d-2}}\right)  -\bar{\gamma}^{\mu}\gamma
^{\nu\lambda}D_{\mu}\varepsilon_{1}=0,\;\varepsilon_{2}\left(  x\right)
=\text{arbitrary.}\label{interm}%
\end{equation}
The equation for $\varepsilon_{1}$ can be manipulated by contracting with
$\bar{\gamma}_{\nu\lambda}$ and using%
\begin{equation}
\bar{\gamma}_{\nu\lambda}\bar{\gamma}^{\nu\lambda}=-d\left(  d-1\right)
,\;\;\bar{\gamma}_{\nu\lambda}\bar{\gamma}^{\mu}\gamma^{\nu\lambda}=-\left(
d-1\right)  \left(  d-4\right)  \bar{\gamma}^{\mu},
\end{equation}
to extract the following expression%
\begin{equation}
\left(  \partial_{\mu}\ln\phi^{\frac{d}{d-2}}\bar{\gamma}^{\mu}\varepsilon
_{1}+\bar{\gamma}^{\mu}D_{\mu}\varepsilon_{1}\;\right)  \left(  d-1\right)
\left(  d-4\right)  =0.\label{intermediate}%
\end{equation}
After some manipulation of gamma matrices Eq.(\ref{interm}) is simplified to
the following form (to verify use $\bar{\gamma}^{\mu}\gamma^{\nu\lambda}%
\gamma_{\mu}=\left(  d-4\right)  \bar{\gamma}^{\nu\lambda}$)
\begin{equation}
D_{\mu}\varepsilon_{1}=\frac{1}{d}\gamma_{\mu}\left(  \bar{\gamma}\cdot
D\varepsilon_{1}\right)  \text{ and }\left(  d-4\right)  \bar{\gamma}^{\mu
}D_{\mu}\left(  \phi^{\frac{d}{d-2}}\varepsilon_{1}\right)  =0,\;\varepsilon
_{2}\left(  x\right)  =\text{arbitrary.}%
\end{equation}
The second equation is trivially satisfied if $d+2=6$ so it is a restriction
on the SO$\left(  d,1\right)  $ spinor $\varepsilon_{1}$ only when
$d+2=5,8,12.$

Now consider flat space as an example, with $\partial_{\mu}\phi=0,$ and
$\omega_{\mu}^{ab}=0.$ The solutions of these equations are
\begin{align}
\varepsilon_{1} &  =\varepsilon_{1}^{0}+x\cdot\gamma\tilde{\varepsilon}%
_{1}^{0},\text{ for }d+2=6,~\text{with }\varepsilon_{1}^{0},\tilde
{\varepsilon}_{1}^{0}\text{ constant spinors of SO}\left(  3,1\right)
\text{.}\label{d4}\\
\varepsilon_{1} &  =\varepsilon_{1}^{0},\;\text{for }d+2=5,8,12,\;\text{with
}\varepsilon_{1}^{0},\text{constant spinor of SO}\left(  d,1\right)  \text{ .}%
\end{align}
Note that in the flat case for $d+2=6,$ the complex SO$\left(  3,1\right)  $
spinors $\varepsilon_{1}^{0},\tilde{\varepsilon}_{1}^{0}$ correspond to
supersymmetry and superconformal transformations respectively, and their
closure gives the superalgebra SU$\left(  2,2|1\right)  .$ On the other hand,
for $d+2=12$ the SO$\left(  9,1\right)  $ spinor $\varepsilon_{1}^{0}$ is real
and contains only 16 components, so this case has only 16 supersymmetries, but
not superconformal symmetry. In a more general curved space $\varepsilon
_{1}\left(  x\right)  $ may depend on $x^{\mu}$ even when $d+2=5,8,12,$ and
may thus contain more than one constant spinor of SO$\left(  d,1\right)  ,$
thus possibly having more than 16 supersymmetries.

The number of supersymmetries may be determined also by analyzing the number
of conserved currents associated with constant spinor parameters. The
conserved current discussed in the text is
\begin{equation}
\bar{\varepsilon}J^{M}=\delta\left(  W\right)  \sqrt{G}\Omega^{\frac{d-4}%
{d-2}}F_{PQ}^{a}V_{N}~\bar{\varepsilon}\left(  \Gamma^{PQN}\bar{\Gamma}%
^{M}\right)  \lambda^{a}.
\end{equation}
Here in general $\bar{\varepsilon}\left(  X\right)  $ depends on the $X^{M}$
in $d+2$ dimensions, while this $\bar{\varepsilon}\left(  X\right)  $
satisfies the SUSY condition (\ref{SUSYcond}). Let's write every component of
$\bar{\varepsilon}J^{M}$ in the $\left(  w,u,x\right)  $ basis$,$ and in the
gauge in which
\begin{align}
\lambda & =\left(
\begin{array}
[c]{c}%
\lambda_{1}\\
0
\end{array}
\right)  e^{\left(  d-1\right)  u},\;F_{wu}=F_{w\mu}=F_{u\mu}=0,\;\Omega=\phi
e^{\left(  d-2\right)  u}\\
\sqrt{G}  & =e^{-2du}\sqrt{-g},\;\Gamma^{w}\sim\left(
\begin{array}
[c]{cc}%
0 & 0\\
1 & 0
\end{array}
\right)  ,\;\Gamma^{u}\sim\left(
\begin{array}
[c]{cc}%
0 & 1\\
0 & 0
\end{array}
\right)
\end{align}
all evaluated at $w=0.$ We get%
\begin{align}
\bar{\varepsilon}J^{w} &  =\frac{1}{2}\delta\left(  w\right)  e^{-5u}\sqrt
{-g}\phi^{\frac{d-4}{d-2}}F_{Pq}^{a}~\bar{\varepsilon}\left(  \Gamma
^{Pq}\Gamma^{w}\bar{\Gamma}^{w}\right)  \left(
\begin{array}
[c]{c}%
\lambda_{1}\\
0
\end{array}
\right)  =0,\\
\bar{\varepsilon}J^{u} &  =\frac{1}{2}\delta\left(  w\right)  e^{-5u}\sqrt
{-g}\phi^{\frac{d-4}{d-2}}F_{Pq}^{a}~\bar{\varepsilon}\left(  \Gamma
^{Pq}\Gamma^{w}\bar{\Gamma}^{u}\right)  \left(
\begin{array}
[c]{c}%
\lambda_{1}\\
0
\end{array}
\right)  =0,\\
\bar{\varepsilon}J^{\mu} &  =\frac{1}{2}\delta\left(  w\right)  e^{-5u}%
\sqrt{-g}\phi^{\frac{d-4}{d-2}}F_{Pq}^{a}~\bar{\varepsilon}\left(  \Gamma
^{Pq}\Gamma^{w}\bar{\Gamma}^{\mu}\right)  \left(
\begin{array}
[c]{c}%
\lambda_{1}\\
0
\end{array}
\right) \\
&  \sim-\frac{1}{2}\delta\left(  w\right)  e^{-5u}\sqrt{-g}\phi^{\frac
{d-4}{d-2}}F_{Pq}^{a}~\left(  \bar{\varepsilon}_{2}\;\bar{\varepsilon}%
_{1}\right)  \left(  \Gamma^{Pq}\bar{\Gamma}^{\mu}\right)  \left(
\begin{array}
[c]{c}%
0\\
\lambda_{1}%
\end{array}
\right) \\
&  =-\frac{1}{2}\delta\left(  w\right)  e^{-5u}\sqrt{-g}\phi^{\frac{d-4}{d-2}%
}F_{Pq}^{a}~\bar{\varepsilon}_{1}\left(  -\bar{\gamma}^{Pq}\gamma^{\mu
}\right)  \lambda_{1}%
\end{align}
So $\varepsilon_{2}$ does not contribute at all in the conformal shadow.
$\varepsilon_{1}$ is a real spinor of SO$\left(  9,1\right)  $ so it has 16
components, implying 16 conserved SUSY currents or 16 supersymmetries if
$\varepsilon_{1}$ is just a constant spinor. Among remaining questions in
12-dimensions is whether there are spacetimes with nontrivial $\bar
{\varepsilon}_{1}\left(  x\right)  $ that contain more than $16$ constant
spinor components (for example, an analog of Eq.(\ref{d4}) in 4-dimensions),
thus implying more than 16 supersymmetries? The fact that the compactified
version of the 12-dimensional theory $\left(  10+2\right)  \rightarrow\left(
4+2\right)  ,$ shown in Fig.1, is symmetric under SU$\left(  2,2|4\right)  $
and contains 32 supersymmetries, is an indication that there may be
non-trivial backgrounds $W,\Omega,G_{MN}$ in which there are at least 32
supersymmetries, but we have not identified them in this paper.

\section{SO$\left(  10,2\right)  $ spinors in SU$\left(  2,2\right)  \times
$SU$\left(  4\right)  $ basis \label{32s}}

We label the SO$\left(  10,2\right)  $ real spinor $\mathbf{32}$ in the
SU$\left(  2,2\right)  \times$SU$\left(  4\right)  $ basis as $\psi_{\alpha
}^{~r},$ which is a complex $\left(  4,\bar{4}\right)  $. The spinor labels
$\rho,r$ in this section should not be confused with the vector labels
$\alpha,m$ used in the previous section. Its conjugate $\bar{\psi}$ will be
labelled as $\bar{\psi}_{r}^{~\rho},$ which is a $\left(  \bar{4},4\right)  $
and is constructed by taking Hermitian conjugation and multiplying by the
SU$\left(  2,2\right)  $ metric in spinor space $\eta^{\dot{\rho}\sigma}$ (see
appendix of \cite{susy2tN1})%
\begin{equation}
\bar{\psi}=\psi^{\dagger}\eta;\;\bar{\psi}_{r}^{~\rho}=\left(  \psi^{\dagger
}\right)  _{r\dot{\sigma}}\eta^{\dot{\sigma}\rho}.
\end{equation}
The charge conjugate spinor $\psi^{c}$ is given by taking the transpose of
$\bar{\psi}$ and multiplying by the charge conjugation matrix
\begin{equation}
\psi^{c}=C\bar{\psi}^{T};\;\;\left(  \psi^{c}\right)  _{\dot{\rho}r}%
=C_{\dot{\rho}\sigma}\left(  \bar{\psi}^{T}\right)  _{~r}^{\sigma}%
=C_{\dot{\rho}\sigma}\left(  \eta^{T}\right)  ^{\sigma\dot{\kappa}}\left(
\psi^{\ast}\right)  _{\dot{\kappa}r}=\tilde{C}_{\dot{\rho}\dot{\kappa}}\left(
\psi^{\ast}\right)  _{\dot{\kappa}r}%
\end{equation}
We define a pseudoreal spinor basis of SO$\left(  10,2\right)  $ that has 32
real components (constructed from the 16 complex components of $\psi$ or
$\psi^{c}$) as follows (here we suppress the Yang-Mills group adjoint
representation label)%
\begin{equation}
\psi_{A}=\frac{1}{\sqrt{2}}\left(
\begin{array}
[c]{c}%
\psi_{\rho}^{~r}\\
\left(  \psi^{c}\right)  _{\dot{\rho}r}%
\end{array}
\right)  \sim32=\left(
\begin{array}
[c]{c}%
\left(  4,\bar{4}\right) \\
\left(  \bar{4},4\right)
\end{array}
\right) \label{lambdapsi}%
\end{equation}
The normalization of $1/\sqrt{2}$ is to insure the correct normalization of
kinetic terms in terms of $\psi.$

The SO$\left(  10,2\right)  $ transformation laws of the 32-spinor $\psi_{A}$%
\begin{equation}
\delta_{\omega}\psi_{A}=-\frac{1}{4}\omega^{MN}\left(  \Gamma_{MN}\right)
_{A}^{~B}\psi_{B},
\end{equation}
can be rewritten in the SO$\left(  4,2\right)  \times$SO$\left(  6\right)
=$SU$\left(  2,2\right)  \times$SU$\left(  4\right)  $ basis as the following
of SO$\left(  10,2\right)  $ transformation laws of $\psi_{\rho}^{~r}%
\sim\left(  4,\bar{4}\right)  $%
\begin{align}
\left(  \delta_{\omega}\psi\right)  _{\rho}^{~r} &  =-\frac{1}{4}\omega
^{mn}\left(  \gamma_{mn}\psi\right)  _{\rho}^{~r}+\frac{1}{4}\omega
^{IJ}\left(  \psi\gamma_{IJ}\right)  _{\rho}^{~r}+\frac{1}{2}\omega
^{mI}\left(  \gamma_{m}\psi^{c}\gamma_{I}\right)  _{\rho}^{~r}\\
&  =-\frac{1}{4}\omega^{mn}\left(  \gamma_{mn}\right)  _{\rho}^{~\sigma}%
\psi_{\sigma}^{~r}+\frac{1}{4}\omega^{IJ}\psi_{\rho}^{~s}\left(  \gamma
_{IJ}\right)  _{s}^{~r}+\frac{1}{2}\omega^{mI}\left(  \gamma_{m}\right)
_{\rho}^{~\dot{\sigma}}\left(  \psi^{c}\right)  _{\dot{\sigma}s}\left(
\bar{\gamma}_{I}\right)  ^{sr}%
\end{align}
Here all small $\gamma^{\prime}s$ are 4$\times4$ matrices expressed in the
spinor bases of SU$\left(  2,2\right)  $ or SU$\left(  4\right)  $. From these
we compute the transformation laws for the charge conjugate spinor $\psi
^{c}\sim\left(  \bar{4},4\right)  $ as
\begin{align}
\left(  \delta_{\omega}\psi^{c}\right)  _{\dot{\rho}r} &  =-\frac{1}{4}%
\omega^{mn}\left(  \bar{\gamma}_{mn}\psi^{c}\right)  _{\dot{\rho}r}+\frac
{1}{4}\omega^{IJ}\left(  \psi^{c}\bar{\gamma}_{IJ}\right)  _{\dot{\rho}%
r}+\frac{1}{2}\omega^{mI}\left(  \bar{\gamma}_{m}\psi\gamma_{I}\right)
_{\dot{\rho}r}\\
&  =-\frac{1}{4}\omega^{mn}\left(  \bar{\gamma}_{mn}\right)  _{\dot{\rho}%
}^{~\dot{\sigma}}\left(  \psi^{c}\right)  _{\dot{\sigma}r}+\frac{1}{4}%
\omega^{IJ}\left(  \psi^{c}\right)  _{\dot{\rho}s}\left(  \bar{\gamma}%
_{IJ}\right)  _{~r}^{s}+\frac{1}{2}\omega^{mI}\left(  \bar{\gamma}_{m}\right)
_{\dot{\rho}}^{~\sigma}\psi_{\sigma}^{~s}\left(  \gamma_{I}\right)  _{sr}%
\end{align}
These are consistent with $\delta_{\omega}\psi^{c}=C\overline{\left(
\delta_{\omega}\psi\right)  }^{T}=C\eta^{T}\left(  \delta_{\omega}\psi\right)
^{\ast}$ since
\begin{gather}
C\eta^{T}\left(  \gamma_{mn}\right)  ^{\ast}\left(  \eta^{T}\right)
^{-1}C^{-1}=\bar{\gamma}_{mn},\;\;C\eta^{T}\left(  \gamma_{m}\right)  ^{\ast
}\left(  \eta^{T}\right)  ^{-1}C^{-1}=-\bar{\gamma}_{m}\\
\left(  \gamma_{IJ}\right)  ^{\ast}=\bar{\gamma}_{IJ},\;\;\left(  \bar{\gamma
}_{I}\right)  ^{\ast}=-\gamma_{I};\text{ also }\left(  \gamma_{I}\right)
_{rs},\left(  \bar{\gamma}_{I}\right)  ^{rs}\text{ are antisymmetric}%
\end{gather}
The last line also implies
\begin{equation}
\left(  \bar{\gamma}_{I}\right)  ^{\dagger}=\gamma_{I},\text{ and }\left(
\gamma_{IJ}\right)  ^{\dagger}=-\gamma_{IJ}\text{ }~\text{,~ }\left(
\bar{\gamma}_{IJ}\right)  ^{\dagger}=-\bar{\gamma}_{IJ}%
\end{equation}
which is consistent with Hermitian SU$\left(  4\right)  $ generators $\frac
{i}{2}\gamma_{IJ}$ and $\frac{i}{2}\bar{\gamma}_{IJ}$ in the $\mathbf{4}$ and
$\mathbf{\bar{4}}$ representations respectively. The explicit matrix form of
the antisymmetric SO$\left(  6\right)  $ gamma matrices $\left(  \gamma
_{I}\right)  _{rs},\left(  \bar{\gamma}_{I}\right)  ^{rs}$ can be taken as
\begin{align}
\left(  \gamma_{I}\right)  _{rs} &  =\left(  \left(  \sigma_{2}\times
i\sigma_{2}\vec{\sigma}\right)  ,\left(  \sigma_{2}\vec{\sigma}\times
\sigma_{2}\right)  \right)  ,\;(\text{note }i\sigma_{2}\vec{\sigma}=\left(
\sigma_{1},i,-\sigma_{3}\right)  \text{)}\label{gamao61}\\
\left(  \bar{\gamma}_{I}\right)  ^{rs} &  =\left(  \left(  \sigma_{2}\times
i\sigma_{2}\vec{\sigma}^{\ast}\right)  ,\left(  -\sigma_{2}\vec{\sigma}^{\ast
}\times\sigma_{2}\right)  \right)  ,\;(\text{note }\left(  -\sigma_{2}%
\vec{\sigma}^{\ast}\right)  =\left(  i\sigma_{3},1,-i\sigma_{1}\right)
\text{)}\label{gammao62}%
\end{align}
These satisfy the Clifford algebra property of SO$\left(  6\right)  $ gamma
matrices%
\begin{equation}
\left(  \gamma_{I}\bar{\gamma}_{J}+\gamma_{J}\bar{\gamma}_{I}\right)
_{r}^{~s}=2\delta_{IJ}\delta_{r}^{~s}.
\end{equation}
Some further property is that for each $I$ these satisfy
\begin{equation}
\left(  \bar{\gamma}_{I}\right)  ^{rs}=\frac{1}{2}\varepsilon^{rsuv}\left(
\gamma_{I}\right)  _{uv}.
\end{equation}

The SO$\left(  10,2\right)  $ transformation laws of the 32-component spinor
$\delta_{\omega}\psi_{A}=-\frac{1}{4}\omega^{mn}\left(  \gamma_{mn}\right)
_{A}^{~B}\psi_{B}$ can now be written in the form of $32\times32$ matrices
\begin{align}
\delta_{\omega}\left(
\begin{array}
[c]{c}%
\psi\\
\psi^{c}%
\end{array}
\right)   &  =-\frac{1}{4}\omega^{MN}\left(  \Gamma_{MN}\right)  \left(
\begin{array}
[c]{c}%
\psi\\
\psi^{c}%
\end{array}
\right) \\
\omega^{MN}\left(  \Gamma_{MN}\right)   &  =\left(
\begin{array}
[c]{cc}%
\omega^{mn}\left(  \gamma_{mn}\otimes1_{4}\right)  +\omega^{IJ}\left(
1_{4}\otimes\gamma_{IJ}\right)  & 2\omega^{mI}\left(  \gamma_{m}\otimes
\bar{\gamma}_{I}\right) \\
2\omega^{mI}\left(  \bar{\gamma}_{m}\otimes\gamma_{I}\right)  & \omega
^{mn}\left(  \bar{\gamma}_{mn}\otimes1_{4}\right)  +\omega^{IJ}\left(
1_{4}\otimes\bar{\gamma}_{IJ}\right)
\end{array}
\right)
\end{align}
The direct products $\otimes$ are applied from (left side) x (right side) on
the 4$\times4$ matrices $\psi,\psi^{c}.$ In this notation the $\overline
{32}\times32$ gamma matrices $\left(  \bar{\Gamma}_{M}\right)  ^{\dot{A}%
B}=\left(  \bar{\Gamma}_{m},\bar{\Gamma}_{I}\right)  $ act such as to mix the
two spinors $\overline{32}$ and $32$ of SO$\left(  10,2\right)  ,$ as
$\bar{\Gamma}_{m}\left(  32\right)  =\left(  \overline{32}\right)  ,$ where%
\begin{equation}
\overline{32}=\left(
\begin{array}
[c]{c}%
\left(  4,4\right) \\
\left(  \bar{4},\bar{4}\right)
\end{array}
\right)  ,~\text{versus}~~32=\left(
\begin{array}
[c]{c}%
\left(  4,\bar{4}\right) \\
\left(  \bar{4},4\right)
\end{array}
\right)  =\left(
\begin{array}
[c]{c}%
\psi\\
\psi^{c}%
\end{array}
\right)
\end{equation}
Therefore, $\left(  \bar{\Gamma}_{M}\right)  ^{\dot{A}B}=\left(  \bar{\Gamma
}_{m},\bar{\Gamma}_{I}\right)  ^{\dot{A}B}$ must act on $\psi_{B}$ as follows
\begin{equation}
\bar{\Gamma}_{m}\left(
\begin{array}
[c]{c}%
\left(  \psi\right)  _{\rho}^{~r}\\
\left(  \psi^{c}\right)  _{\dot{\rho}r}%
\end{array}
\right)  =\left(
\begin{array}
[c]{c}%
\left(  \gamma_{m}\psi^{c}\right)  _{\rho r}\\
\left(  \bar{\gamma}_{m}\psi\right)  _{\dot{\rho}}^{~r}%
\end{array}
\right)  \sim\overline{32},\;\text{and }\;\bar{\Gamma}_{I}\left(
\begin{array}
[c]{c}%
\left(  \psi\right)  _{\rho}^{~r}\\
\left(  \psi^{c}\right)  _{\dot{\rho}r}%
\end{array}
\right)  =\left(
\begin{array}
[c]{c}%
\left(  \psi\gamma_{I}\right)  _{\rho r}\\
-\left(  \psi^{c}\bar{\gamma}_{I}\right)  _{\dot{\rho}}^{r}%
\end{array}
\right)  \sim\overline{32}%
\end{equation}
Now we can introduce the SO$\left(  10,2\right)  $ gamma matrices $\Gamma_{M}$
and $\bar{\Gamma}_{M}$ as follows%
\begin{align}
\overline{32}\times32,\;\left(  \bar{\Gamma}_{M}\right)  ^{\dot{A}B} &
:\;\;\bar{\Gamma}_{m}=\left(
\begin{array}
[c]{cc}%
0 & \gamma_{m}\otimes1_{4}\\
\bar{\gamma}_{m}\otimes1_{4} & 0
\end{array}
\right)  ,\;\bar{\Gamma}_{I}=\left(
\begin{array}
[c]{cc}%
1_{4}\otimes\gamma_{I} & 0\\
0 & -1_{4}\otimes\bar{\gamma}_{I}%
\end{array}
\right) \\
32\times\overline{32},\;\left(  \Gamma_{M}\right)  _{A\dot{B}} &
:\;\;\Gamma_{m}=\left(
\begin{array}
[c]{cc}%
0 & \gamma_{M}\otimes1_{4}\\
\bar{\gamma}_{m}\otimes1_{4} & 0
\end{array}
\right)  ,\;\Gamma_{I}=\left(
\begin{array}
[c]{cc}%
1_{4}\otimes\bar{\gamma}_{I} & 0\\
0 & -1_{4}\otimes\gamma_{I}%
\end{array}
\right)
\end{align}
These SO$\left(  10,2\right)  $ gamma matrices $\Gamma_{M}=\left(  \Gamma
_{m},\Gamma_{I}\right)  ,$ and $\bar{\Gamma}_{M}=\left(  \bar{\Gamma}_{m}%
,\bar{\Gamma}_{I}\right)  $ satisfy the Clifford algebra property
\begin{equation}
\Gamma_{M}\bar{\Gamma}_{N}+\Gamma_{N}\bar{\Gamma}_{M}=2\eta_{MN},\text{ and
\ }\bar{\Gamma}_{M}\Gamma_{N}+\bar{\Gamma}_{N}\Gamma_{N}=2\eta_{MN}%
\end{equation}
In more detail, this is seen as follows
\begin{align}
\Gamma_{m}\bar{\Gamma}_{n}+\Gamma_{n}\bar{\Gamma}_{m} &  =\left(
\begin{array}
[c]{cc}%
0 & \gamma_{m}\otimes1_{4}\\
\bar{\gamma}_{m}\otimes1_{4} & 0
\end{array}
\right)  \left(
\begin{array}
[c]{cc}%
0 & \gamma_{n}\otimes1_{4}\\
\bar{\gamma}_{n}\otimes1_{4} & 0
\end{array}
\right)  +\left(  m\leftrightarrow n\right) \\
&  =\left(
\begin{array}
[c]{cc}%
\left(  \gamma_{m}\bar{\gamma}_{n}+\left(  m\leftrightarrow n\right)  \right)
\otimes1_{4} & 0\\
0 & \left(  \bar{\gamma}_{m}\gamma_{n}+\left(  m\leftrightarrow n\right)
\right)  \otimes1_{4}%
\end{array}
\right)  =2\eta_{mn}%
\end{align}
Similarly,%
\begin{align}
\Gamma_{I}\bar{\Gamma}_{J}+\Gamma_{J}\bar{\Gamma}_{I} &  =\left(
\begin{array}
[c]{cc}%
1_{4}\otimes\bar{\gamma}_{I} & 0\\
0 & -1_{4}\otimes\gamma_{I}%
\end{array}
\right)  \left(
\begin{array}
[c]{cc}%
1_{4}\otimes\gamma_{J} & 0\\
0 & -1_{4}\otimes\bar{\gamma}_{J}%
\end{array}
\right)  +\left(  I\leftrightarrow J\right) \\
&  =\left(
\begin{array}
[c]{cc}%
1_{4}\otimes\left(  \gamma_{J}\bar{\gamma}_{I}+\left(  I\leftrightarrow
J\right)  \right)  & 0\\
0 & 1_{4}\otimes\left(  \bar{\gamma}_{J}\gamma_{I}+\left(  I\leftrightarrow
J\right)  \right)
\end{array}
\right)  =2\delta_{IJ}%
\end{align}
Note that in computing the products above the orders in the second factor are
reversed because the second factor in the direct product is applied from the
\textit{right side} as emphasized above. Finally,%
\begin{align}
\Gamma_{m}\bar{\Gamma}_{I}+\Gamma_{I}\bar{\Gamma}_{m} &  =\left(
\begin{array}
[c]{cc}%
0 & \gamma_{m}\otimes1_{4}\\
\bar{\gamma}_{m}\otimes1_{4} & 0
\end{array}
\right)  \left(
\begin{array}
[c]{cc}%
1_{4}\otimes\gamma_{I} & 0\\
0 & -1_{4}\otimes\bar{\gamma}_{I}%
\end{array}
\right) \\
&  +\left(
\begin{array}
[c]{cc}%
1_{4}\otimes\bar{\gamma}_{I} & 0\\
0 & -1_{4}\otimes\gamma_{I}%
\end{array}
\right)  \left(
\begin{array}
[c]{cc}%
0 & \gamma_{m}\otimes1_{4}\\
\bar{\gamma}_{m}\otimes1_{4} & 0
\end{array}
\right) \\
&  =\left(
\begin{array}
[c]{cc}%
0 & -\left(  \gamma_{m}\otimes\bar{\gamma}_{I}\right)  +\left(  \gamma
_{m}\otimes\bar{\gamma}_{I}\right) \\
\left(  \bar{\gamma}_{m}\otimes\gamma_{I}\right)  -\left(  \bar{\gamma}%
_{m}\otimes\gamma_{I}\right)  & 0
\end{array}
\right) \\
&  =0
\end{align}

In the same way we compute $\left(  \Gamma_{MN}\right)  _{A}^{~B}=\frac{1}%
{2}\left(  \Gamma_{M}\bar{\Gamma}_{N}-\Gamma_{N}\bar{\Gamma}_{M}\right)
_{A}^{~B}$ and find%
\begin{align}
\Gamma_{mn} &  =\left(
\begin{array}
[c]{cc}%
\gamma_{mn}\otimes1_{4} & 0\\
0 & \bar{\gamma}_{mn}\otimes1_{4}%
\end{array}
\right)  ,\;\Gamma_{IJ}=\left(
\begin{array}
[c]{cc}%
-1_{4}\otimes\gamma_{IJ} & 0\\
0 & -1_{4}\otimes\bar{\gamma}_{IJ}%
\end{array}
\right)  ,\;\\
\Gamma_{mI} &  =\left(
\begin{array}
[c]{cc}%
0 & -\gamma_{m}\otimes\bar{\gamma}_{I}\\
\bar{\gamma}_{m}\otimes\gamma_{I} & 0
\end{array}
\right)
\end{align}
The matrices $\frac{1}{2i}\gamma_{mn}$ close under commutation to form the
32$\times32$ spinor representation of the SO$\left(  10,2\right)  $ Lie algebra.

There is an antisymmetric SO$\left(  10,2\right)  $ invariant tensor
$a_{AB}=-a_{BA}$ in the space of the spinors since $\left(  32\times32\right)
_{antisymm}$ contains the SO$\left(  10,2\right)  $ singlet, namely the matrix
$a$ satisfies $\delta_{\omega}a=0,$ or $\Gamma_{MN}a+a\left(  \Gamma
_{MN}\right)  ^{T}=0.$ Taking the antisymmetry of $a$ into account, this
implies that the matrices $\left(  \gamma_{mn}a\right)  _{AB}$ are symmetric
under the interchange of $A\leftrightarrow B,$ $\left(  \Gamma_{MN}a\right)
=\left(  \Gamma_{MN}a\right)  ^{T}.$ The explicit matrix $a$ is given by%
\begin{equation}
a_{AB}=\left(
\begin{array}
[c]{cc}%
0 & C\otimes1_{4}\\
\bar{C}\otimes1_{4} & 0
\end{array}
\right)  .
\end{equation}
Recalling the following symmetry properties of the gamma matrices under
transposition (appendix in \cite{susy2tN1})%
\begin{align}
\left(  \Gamma^{M}\bar{C}\right)  ^{T} &  =-\left(  \Gamma^{M}\bar{C}\right)
,\;\left(  \Gamma^{MN}C\right)  ^{T}=\left(  \bar{\Gamma}^{MN}C\right)  ,\\
\left(  \Gamma^{I}\right)  ^{T} &  =-\left(  \Gamma^{I}\right)  ,\;\left(
\Gamma^{IJ}\right)  ^{T}=\left(  \bar{\Gamma}^{IJ}\right)  ,
\end{align}
we verify explicitly that indeed $\left(  \Gamma_{MN}a\right)  =\left(
\Gamma_{MN}a\right)  ^{T}$ is satisfied. The matrix $a_{AB},$ together with
its inverse $a^{AB},$ plays the role of an invariant metric that can be used
to raise or lower indices in the 32-spinor space.

$\allowbreak$

\begin{acknowledgments}
We gratefully acknowledge discussions with S.H. Chen. I. Bars thanks the CERN Theory Division for hospitality and for providing a stimulating atmosphere while this research was completed.
\end{acknowledgments}


\begin{thebibliography}{99}                                                                                               %
\bibitem {2tstandardM}I. Bars, \textquotedblleft The standard model of
particles and forces in the framework of 2T-physics\textquotedblright, Phys.
Rev. \textbf{D74} (2006) 085019 [arXiv:hep-th/0606045]. For a summary see
\textquotedblleft The Standard Model as a 2T-physics theory\textquotedblright,
Proc. of SUSY06: 14th Int. Conference on Supersymmetry and the Unification of
Fundamental Interactions, Irvine, California, 12-17 Jun 2006,\ arXiv:hep-th/0610187.

\bibitem {susy2tN1}I. Bars and Y.C. Kuo, \textquotedblleft\textquotedblleft
Field theory in 2T-physics with $N=1$ supersymmetry\textquotedblright\ Phys.
Rev. Lett. \textbf{99} (2007) 41801 [arXiv:hep-th/0703002]; ibid.
\textquotedblleft Supersymmetric field theory in 2T-physics,\textquotedblright%
\ Phys. Rev. \textbf{D76 }(2007) 105028,. [arXiv:hep-th/0703002].

\bibitem {susy2tN2N4}I. Bars and Y.C. Kuo, \textquotedblleft\textquotedblleft%
\ N=2,4 Supersymmetric Gauge Field Theory in 2T-physics\textquotedblright%
\ Phys. Rev. \textbf{D79} (2009) 025001 [arXiv:0808.0537].

\bibitem {emergentfieldth1}I. Bars, S-H. Chen and G. Quelin, \textquotedblleft
Dual Field Theories in (d-1)+1 Emergent Spacetimes from a Unifying Field
Theory in d+2 Spacetime,\textquotedblright\ Phys. Rev. \textbf{D76} (2007)
065016 [arXiv:0705.2834 [hep-th]].

\bibitem {emergentfieldth2}I. Bars, and G. Quelin, \textquotedblleft Dualities
among 1T-Field Theories with Spin, Emerging from a Unifying 2T-Field
Theory\textquotedblright, Phys. Rev. \textbf{D77} (2008) 125019
[arXiv:0802.1947 [hep-th]].

\bibitem {2tGravity}I. Bars, \textquotedblleft Gravity in
2T-Physics\textquotedblright, Phys. Rev. \textbf{D77} (2008) 125027
[arXiv:0804.1585 [hep-th]].

\bibitem {2tGravDetails}I. Bars and S-H Chen, \textquotedblleft Geometry and
Symmetry Structures in 2T Gravity\textquotedblright, Phys. Rev. D79 (2009)
085021 [arXiv:0811.2510 (hep-th)].

\bibitem {2tCosmo}I. Bars and S-H Chen, \textquotedblleft\ The Big Bang and
Inflation United by an Analytic Solution," [arXiv:1004.0752] to appear in
Phys. Rev D.

\bibitem {phaseSpace}I. Bars, \textquotedblleft Gauge Symmetry in Phase Space,
Consequences for Physics and Spacetime,\textquotedblright\ [arXiv:1004.0688
(hep-th)], to appear in the International Journal of Modern Physics A (IJMPA).

\bibitem {2tScalars}I. Bars, \textquotedblleft Constraints on Interacting
Scalars in 2T Field Theory and No Scale Models in 1T Field
Theory\textquotedblright, arXiv:1008.1540 [hep-th], submitted to Phys. Rev. D.

\bibitem {2tsugra}I. Bars, \textquotedblleft2T-Supergravity\textquotedblright,
in preparation.

\bibitem {2tbacgrounds}I. Bars, \textquotedblleft Two time physics with
gravitational and gauge field backgrounds", Phys. Rev. \textbf{D62}, 085015
(2000) [arXiv:hep-th/0002140]; I. Bars and C. Deliduman, \textquotedblleft%
\ High spin gauge fields and two time physics", Phys. Rev. \textbf{D64},
045004 (2001) [arXiv:hep-th/0103042].

\bibitem {Dirac}P.A.M Dirac, Ann. Math. \textbf{37} (1936) 429.

\bibitem {kastrup}H. A. Kastrup, Phys. Rev. \textbf{150} (1966) 1183.

\bibitem {salam}G. Mack and A. Salam, Ann. Phys. \textbf{53} (1969) 174.

\bibitem {adler}S. Adler, Phys. Rev. \textbf{D6} (1972) 3445; \textit{ibid}.
\textbf{D8} (1973) 2400.

\bibitem {ferrara}S. Ferrara, A. F. Grillo, and R. Gatto, Ann. Phys. (NY) 76
(1973) 161; S. Ferrara, Nucl. Phys. \textbf{B77} (1974) 73.

\bibitem {fronsdal}F. Bayen, M. Flato, C. Fronsdal and A. Haidari, Phys. Rev.
\textbf{D32} (1985) 2673.

\bibitem {siegel}W. Siegel, Int. J. Mod. Phys. \textbf{A3} (1988) 2713; Int.
Jour. Mod. Phys. \textbf{A4} (1989) 2015.

\bibitem {vasiliev}C. R. Preitschopf and M. A. Vasiliev, Nucl. Phys.
\textbf{B549} (1999) 450 [arXiv:hep-th/9812113].

\bibitem {vasiliev2}M. A. Vasiliev, JHEP \textbf{12} (2004) 046, [hep-th/0404124].

\bibitem {marnelius}R. Marnelius, Phys. Rev. D20, 2091 (1979); R. Marnelius
and B. Nilsson, Phys. Rev. \textbf{D22} (1980) 830; P. Arvidsson and R.
Marnelius, \textquotedblleft Conformal theories including conformal gravity as
gauge theories on the hypercone\textquotedblright\ [arXiv:hep-th/0612060].

\bibitem {weinberg}S. Weinberg, \textquotedblleft Six-dimensional Methods for
Four-dimensional Conformal Field Theories\textquotedblright, arXiv:1006.3480 [hep-th].

\bibitem {rivelles}J.E. Frederico and V.O. Rivelles, \textquotedblleft The
Transition Amplitude for 2T Physics\textquotedblright, arXiv:1002.1263 [hep-th].

\bibitem {AreaDiff}I. Bars, \textquotedblleft Strings from Reduced Large N
Gauge Theory via Area Preserving Diffeomorphisms\textquotedblright, Phys.
Lett. \textbf{B245} (1990) 35.

\bibitem {matrixTh1}T. Banks, W. Fischler, S.H. Shenker, L.
Susskind,\textquotedblleft M Theory As A Matrix Model: A
Conjecture\textquotedblright, Phys. Rev. \textbf{D55} (1997) 5112
[arXiv:hep-th/9610043]; ibid. \textquotedblleft Instantons, Scale Invariance
and Lorentz Invariance in Matrix Theory\textquotedblright, Phys. Lett.
\textbf{B408} (1997) 111 [arXiv:hep-th/9705190]

\bibitem {matrixTh2}T. Banks, \textquotedblleft Matrix
Theory\textquotedblright, arXiv:hep-th/9710231.

\bibitem {matrixThJ1}N. Ishibashi, H. Kawai, Y. Kitazawa and A. Tsuchiya,
Nucl. Phys. B498 (1997), 467 [arXiv:hep-th/9612115].

\bibitem {matrixThJ2}M. Fukuma, H. Kawai. Y. Kitazawa and A. Tsuchiya, Nucl.
Phys. B510 (1998), 158 [arXiv:hep-th/9705128].

\bibitem {matrixThJ3}H. Aoki, S. Iso, H. Kawai, Y. Kitazawa and T. Tada,
\textquotedblleft Space-Time Structures from IIB Matrix
Model\textquotedblright, Prog. Theor. Phys. 99 (1998), 713 [hep-th/9802085].

\bibitem {matrixThJ4}H. Aoki, S. Iso, H. Kawai, Y. Kitazawa, T. Tada, A.
Tsuchiya, \textquotedblleft IIB Matrix Model\textquotedblright, Prog. Theor.
Phys. Suppl. \textbf{134} (1999) 47 [arXiv:hep-th/9908038].

\bibitem {GSW}M.B. Green, J.H. Schwarz and E. Witten, \textit{Superstring
Theory, Volume 2, }Cambridge University Press 1987.

\bibitem {ftheory}C. Vafa, \textquotedblleft F-theory\textquotedblright, Nucl.
Phys. \textbf{B469} (1996) 403, [arXiv:hep-th/9602022].

\bibitem {Stheory}I. Bars, \textquotedblleft Duality and hidden
dimensions\textquotedblright, Lecture in 1995 conference, appeared in
\textit{Frontiers in quantum field theory}, Ed. H. Itoyonaka, M. Kaku. H.
Kunitomo, M. Ninomiya, H. Shirokura, Singapore, World Scientific, 1996, page
52 [hep-th/9604200]; I. Bars, \textquotedblleft Supersymmetry, p-brane duality
and hidden space and time dimensions,\textquotedblright\ Phys. Rev.
\textbf{D54} (1996) 5203, [hep-th/9604139]; I. Bars, \textquotedblleft
S-theory\textquotedblright, Phys. Rev. \textbf{D55} (1997) 2373
[hep-th/9607112]; I. Bars, \textquotedblleft Algebraic Structures in
S-Theory\textquotedblright, talk at conferences Strings-96 and 2$^{nd}$
Sakharov conference, hep-th/9608061.

\bibitem {noncommutative}I. Bars, \textquotedblleft u$_{\ast}\left(
1,1\right)  $ non-commutative gauge theory as the foundation of 2T-physics in
field theory\textquotedblright, Phys. Rev. D64 (2001) 126001 [hep-th/0106013].
I. Bars and S. Rey, \textquotedblleft Noncommutative Sp(2,R) gauge theories: A
Field theory approach to two time physics.\textquotedblright, Phys. Rev. D64
(2001) 046005 [hep-th/0104135].
\end{thebibliography}
\end{document}